\documentclass[a4paper,11pt]{article}
\pdfoutput=1 % if your are submitting a pdflatex (i.e. if you have
\usepackage{jcappub} % for details on the use of the package, please
\usepackage[T1]{fontenc} % if needed
\usepackage{graphicx}
\usepackage[greek, english]{babel}
\usepackage{amsmath, amssymb, mathtools, slashed, physics, xcolor, caption, wrapfig}
\usepackage{threeparttable}
\usepackage{tikz, environ}
\usetikzlibrary{shapes.symbols, shapes, arrows, positioning, backgrounds, fit, er}
\renewcommand{\d}[1]{\ensuremath{\operatorname{d}\!{#1}}}

\newcommand{\code}{ATHE$\nu$A}
\title{Neutrino and pair creation in reconnection-powered coronae of accreting black holes}

\author[1]{D. Karavola,}
\author[1, 2]{M. Petropoulou,}
\author[3]{D. F.~G.~Fiorillo,}
\author[4]{L. Comisso,}
\author[4,5]{L. Sironi}

\affiliation[1]{National and Kapodistrian University of Athens, University Campus Zografos, GR 15784, Athens, Greece}
\affiliation[2]{Institute of Accelerating Systems \& Applications, University Campus Zografos, GR 15784, Athens, Greece}
\affiliation[3]{Deutsches Elektronen-Synchrotron DESY,
Platanenallee 6, 15738 Zeuthen, Germany}
\affiliation[4]{Department of Astronomy and Columbia Astrophysics 
Laboratory, Columbia University, New York, NY 10027, USA}
\affiliation[5]{Center for Computational Astrophysics, Flatiron Institute,
New York, NY, 10010, USA}

\emailAdd{dkaravola@phys.uoa.gr}
\emailAdd{mpetropo@phys.uoa.gr}
\emailAdd{damiano.fiorillo@desy.de}
\emailAdd{luca.comisso@columbia.edu}
\emailAdd{lsironi@astro.columbia.edu}

\abstract{A ubiquitous feature of accreting black hole systems is their hard X-ray emission which is thought to be produced through Comptonization of soft photons by electrons and positrons in the vicinity of the black hole, in a region with optical depth of order unity. The origin and composition of this Comptonizing region, known as the corona, is a matter open for debate. In this paper we investigate the role of relativistic protons accelerated in black-hole magnetospheric current sheets for the pair enrichment and neutrino emission of AGN coronae. Our model has two free parameters, namely the proton plasma magnetization $\sigma_{\rm p}$, which controls the peak energy of the neutrino spectrum, and the Eddington ratio $\lambda_{\rm X, Edd}$ (defined as the ratio between X-ray luminosity $L_{\rm X}$ and Eddington luminosity $L_{\rm Edd}$), which controls the amount of energy transferred to secondary particles. For sources with $\lambda_{\rm X, Edd} \gtrsim \lambda_{\rm Edd, crit}$ (where $\lambda_{\rm Edd, crit} \sim 10^{-1}$ for $\sigma_{\rm p}=10^5$ or $\sim 10^{-2}$ for $\sigma_{\rm p}=10^7$), proton-photon interactions and $ \gamma \gamma$ annihilation produce enough secondary pairs to achieve Thomson optical depths  $\tau_{\rm T} \sim 0.1-10$. In the opposite case of $\lambda_{\rm X, Edd} \lesssim \lambda_{\rm Edd, crit}$, the coronal pairs cannot originate only from hadronic interactions. Additionally, we find that the neutrino luminosity scales as $L^2_{\rm X}/L_{\rm Edd}$ for $\lambda_{\rm X, Edd} \lesssim \lambda_{\rm Edd, crit}$, while it is proportional to $L_{\rm X}$ for higher $\lambda_{\rm X, Edd}$ values. We apply our model to four Seyfert galaxies, including NGC 1068, and discuss our results in light of recent IceCube observations. }

\begin{document}
\maketitle
\flushbottom

\section{Introduction} \label{sec:intro}

Active Galactic Nuclei (AGN) are the most powerful steady emitters of non-thermal radiation. They are likely powered by accretion onto a central supermassive black hole. The accretion disk feeding the black hole is responsible for the emission of optical-ultraviolet (OUV) radiation, while at higher energies from keV to hundreds of keV, in the X-ray band, usually a power-law emission is observed; a typical AGN spectrum is shown, e.g., in figure 1 of Ref.~\cite{2017MNRAS.465..358C}. The origin of the X-rays is traditionally attributed to a compact region close to the black hole, the so-called corona, where a population of hot electrons can Comptonize soft OUV photons, leading to the observed power law~\cite{1980A&A....86..121S, 1991ApJ...380L..51H}. From the point of view of energetics, this moves the question to what is the mechanism that keeps heating the electrons in the corona, as they would otherwise cool due to strong radiative losses.

One possible mechanism for maintaining the high temperature of electrons in the corona is magnetic reconnection, which has also been proposed for stellar-mass accreting black holes (e.g. \cite{1999ApJ...510L.123B, 2002ApJ...572L.173L}). Recently, Ref.~\cite{2017ApJ...850..141B} pointed out that if current sheets form in accreting systems with high density of seed photons, inverse Compton cooling would keep the electron population at low temperatures ($ kT_{\rm e} \ll 100$~keV). 
These ``cold'' electrons are found in plasmoids that move with mildly relativistic speeds and are formed in current sheets. 
Interestingly, radiative particle-in-cell (PIC) simulations have shown that the bulk motions of cold plasma in the current sheet can mimic a quasi-thermal electron distribution with an effective temperature close to 100 keV \cite{2020ApJ...899...52S, 2021MNRAS.507.5625S, sridhar_comptonization_2022}. Additionally, a large fraction  of the dissipated magnetic energy is converted into X-ray radiation. In this context, the magnetic field strength in the coronal region can be inferred by the observed X-ray flux.  

In the last few years, the question of coronal AGN emission has become of new relevance \cite{murase_hidden_2020, kheirandish_high-energy_2021, murase_high-energy_2023}, after the detection of high-energy neutrinos from NGC 1068~\cite{2022Sci...378..538I, inoue_origin_2020, eichmann_solving_2022}, a Seyfert galaxy at $\simeq 10.1$~Mpc. The integrated all-flavor neutrino luminosity is $ L_{\rm \nu + \bar{\nu}} \simeq 10^{42}$~erg/s, while a comparable TeV $ \gamma$-ray emission is not observed \cite{MAGIC-UL-NGC1068}. This suggests as a site for neutrino emission a compact region where $ \gamma$-rays would be attenuated, so that the corona emerges as a natural candidate~\cite{2022ApJ...941L..17M}. This raises new questions on the nature of the non-thermal processes powering the corona, since not only leptons but also hadrons need to be energized to power neutrino emission.

\begin{figure}
        \centering
        \includegraphics[width=0.55\textwidth]{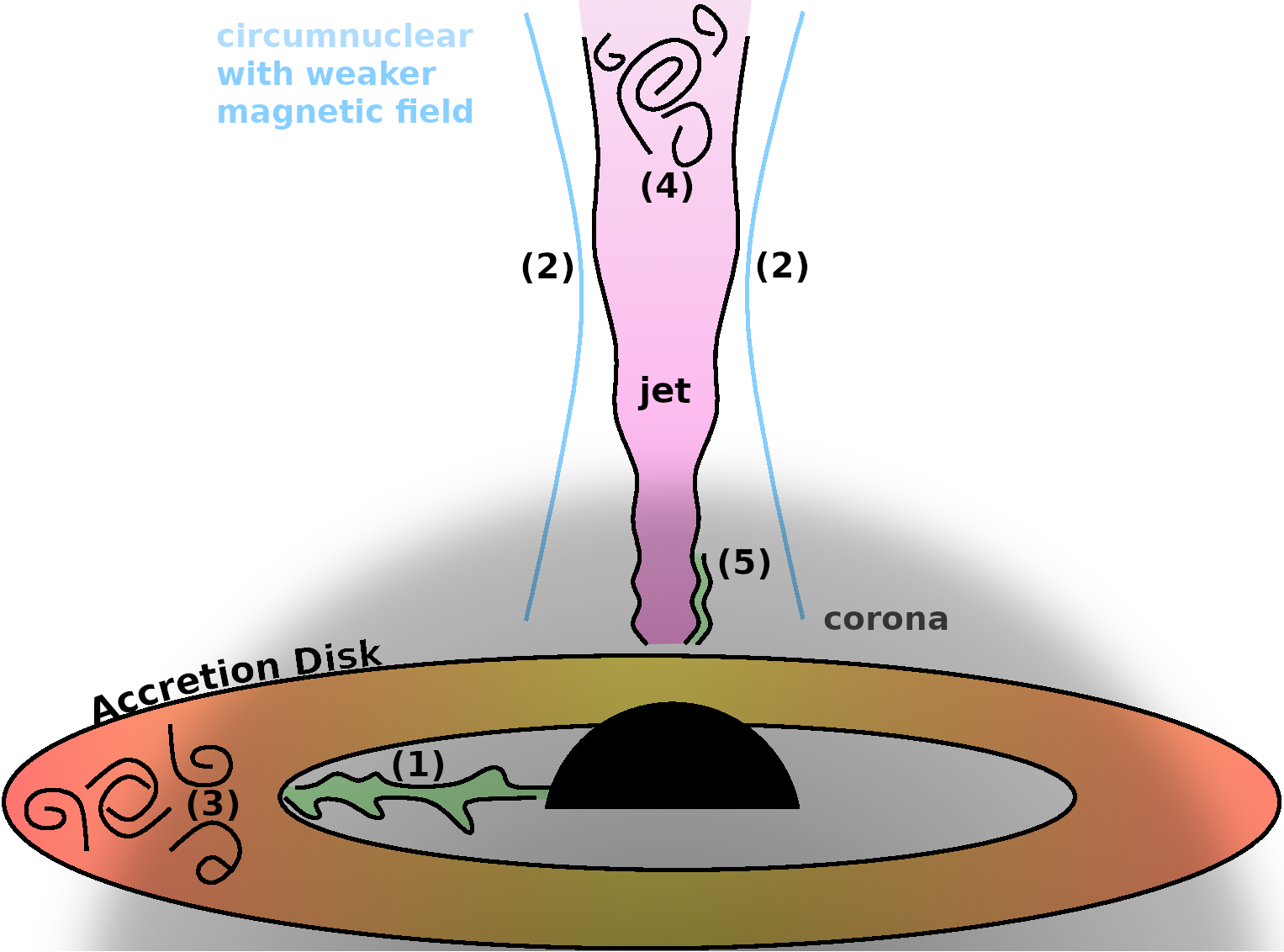}
        \caption{AGN sketch showing the central massive black hole surrounded by an accretion disk. Perpendicularly to the disk a relativistic jet can be formed. A corona can be shown close to the central object. The numbers in the sketch refer to: (1) magnetospheric current sheet, (2) circumnuclear medium, (3) turbulent accretion disk (4) jet and (5) jet boundary current sheet.}
        \label{fig:AGN_sketch}
\end{figure}

A scenario which incorporates magnetic reconnection powering hadron acceleration was recently proposed by some of us in Ref.~\cite{2024ApJ...961L..14F}. In this scenario, a large-scale reconnection layer forming in the black hole magnetospheric region (see sketch in figure~\ref{fig:AGN_sketch}) accelerates protons to energies of tens of TeV. The subsequent $ p\gamma$ interactions with the coronal X-ray photons trigger an electromagnetic cascade. The latter results in a production of pairs maintaining optical thickness (Thomson optical depth of the corona $\gtrsim$1), while also powering a neutrino signal consistent with the IceCube observations. While this study was motivated by the IceCube results of NGC~1068, and as such it was tailored to this source, it remains an open question to understand what features are likely generic to many AGN coronae, and how do the expectations for the neutrino signal vary among different sources.

In this work, we address this question, by conducting a comprehensive study of the neutrino emission and pair density expected from the coronal region of AGN in the presence of magnetospheric current sheets. The predictions for the electromagnetic and neutrino signal depend on only three parameters, namely the black hole mass, the Eddington ratio of the accreting system (defined using the X-ray luminosity of the corona), and the proton plasma magnetization. We perform numerical calculations of the electromagnetic cascade and hadronic neutrino emission, comparing the results with the analytical expectations, to provide the scalings of the expected signal with these three parameters. In this way, we are able to make predictions on the expected neutrino signal from other AGN cores similar to NGC~1068.

This paper is structured as follows. In Section \ref{sec:model} we present a detailed description of our coronal model, and in Section~\ref{sec:analytic} we provide analytical expressions for the predicted neutrino luminosity and secondary pair density. In Section \ref{sec:res} we present the numerical results of our parameter study using the code \code . In Section \ref{sec:seyferts} we test our model on individual Seyfert galaxies and compare the numerical results for the neutrino spectra to those observed by IceCube. We conclude in Section \ref{sec:discuss} with a summary of our results and a discussion.

\section{Model}\label{sec:model}

We set up a model describing the coronal environment in which the X-ray emission of the corona is powered by magnetic dissipation at a large-scale magnetic reconnection layer with a typical size $R$ of a few gravitational radii \cite{Ripperda_2020, el_mellah_reconnection-driven_2023, nathanail_magnetic_2022, sironi_relativistic_2015}. In the reconnection layer the non-thermal particles receive a significant fraction of the magnetic field energy. If the aforementioned particles were not to cool, their energy density would be in rough equipartition with the magnetic field
%, as they receive a fraction of the dissipated magnetic energy
\citep[e.g.][]{sironi_relativistic_2015, 2019ApJ...880...37P}. Motivated by theory and simulations \cite{2017ApJ...850..141B, 2020ApJ...899...52S}, we express the broadband X-ray luminosity of the corona $L_{\rm X}$ as 
    \begin{equation}
        L_{\rm X}= \eta_{\rm X} S_{\rm P} A = \eta_{\rm X} \frac{c}{4 \pi} E_{\rm rec} B A
        \label{eq:Lx}
    \end{equation} 
where $\eta_{\rm X} \sim 0.5$ is the fraction of the magnetic energy dissipated to the X-ray photon field, $S_{\rm P}$ is the Poynting energy flux, $A$ is the surface of the reconnection layer (taken to be $4 \pi R^2$), $B$ and $E_{\rm rec}=\beta_{\rm rec} B$ are respectively the magnetic field and electric field strengths in the upstream region of the layer, with $\beta_{\rm rec} \sim 0.1$ being the reconnection speed in units of the speed of light $c$ \cite{chernoglazov_high-energy_2023, zhang_origin_2023, zhang_fast_2021}. Then, the magnetic field can be expressed as
    \begin{equation}
        B = \left(\frac{L_{\rm X}}{\eta_{\rm X} c \beta_{\rm rec} R^2} \right)^{\frac{1}{2}}. \label{eq:B_field}
    \end{equation} 
We assume that X-ray photons are produced via Comptonization of low-energy seed photons (e.g. from the accretion disk) by electron-positron pairs in the corona that do not necessarily have a hadronic origin. We do not attempt to model in detail the spectral formation of the X-ray coronal emission, but we discuss potential improvements in this regard in Section~\ref{sec:discuss}.  Instead, we adopt an observationally motivated X-ray photon spectrum, in this regard a power law with photon index $s_{\rm X} \sim 2$ extending from $E_{\rm X,\min}=0.1$~keV to $E_{\rm X,\max}=100$~keV,
    \begin{equation}
    \begin{array}{cc}
         n_{\rm X}(E)=n_{\rm X,0}E^{-s_{\rm X}}, & E_{\rm X,\min} \le E  \le E_{\rm X,\max},
    \end{array} 
    \end{equation}
 where $n_{\rm X}(E)$ is the differential number density of photons in the corona. The X-ray spectra of AGN suggest that the optical depth for Comptonization is moderate (i.e., $\tau_{\rm T} \sim 0.1-10$, \cite{1991ApJ...380L..51H, 2018MNRAS.480.1819R, 2020A&A...634A..85P}). We can therefore estimate the pair density that is needed to produce the assumed X-ray spectrum, as $n_{\rm \pm} = \tau_{\rm T}/(\sigma_{\rm T} R)$, where $\sigma_{\rm T}\simeq 6.65 \cdot 10^{-25}$~cm$^2$ is the Thomson cross section. The magnetization of the pair plasma is then computed as 
 \begin{equation}
     \sigma_e = \frac{B^2}{4 \pi n_{\rm \pm} m_{\rm e} c^2} = \frac{\sigma_{\rm T} L_{\rm X}}{4 \pi \eta_{\rm X} \beta_{\rm rec} \tau_{\rm T} R  m_{\rm e} c^3} = \frac{\ell_{\rm X}}{ \eta_{\rm X} \beta_{\rm rec} \tau_{\rm T}}, 
     \label{eq:sigma_e}
 \end{equation}
where $\ell_{\rm X}=\sigma_{\rm T} L_{\rm X}/4\pi R m_{\rm e} c^3$ is the X-ray compactness of the corona. The latter can also be expressed in terms of the Eddington luminosity of an accreting black hole with mass $M_{\rm bh}$, $L_{\rm Edd} \simeq 1.3 \cdot 10^{45}  M_{\rm bh}/(10^7 M_{\odot})~\rm erg~s^{-1}$,  as: 
\begin{equation}
    \ell_{\rm X} = \frac{L_{\rm X}}{L_{\rm Edd}}\frac{1}{\tilde{R}}\frac{m_{\rm p}}{m_{\rm e}}  = \lambda_{\rm X, Edd}\frac{1}{\tilde{R}} \frac{m_{\rm p}}{m_{\rm e}},\label{eq:compact} 
\end{equation}
where $\tilde{R}=R/R_{\rm g}$, $R_{\rm g} =GM_{\rm bh}/c^2$ is the black-hole gravitational radius, and $\lambda_{\rm X, Edd}$ is the Eddington ratio (defined using the broadband X-ray luminosity and not the bolometric luminosity as commonly done in AGN studies). The expected pair magnetization of the coronal plasma is estimated as 
\begin{equation}
    \sigma_{\rm e} \approx 3700 \, \lambda_{\rm X, Edd, -2} \tau^{-1}_{\rm T,-1} \eta^{-1}_{\rm X, -0.3} \beta^{-1}_{\rm rec, -1}\tilde{R}^{-1}
    \label{eq:sigma_e_Edd}
\end{equation}
where we introduced the notation $q_x = q/10^x$.

Driven by PIC simulation results of relativistic magnetic reconnection ($\sigma_e \gg 1$) we expect that any protons, if present in the upstream region of the layer, will be injected to the acceleration process, and energize at a rate ${\rm d}E_{\rm p}/{\rm d}t = e\beta_{\rm rec} B c$, where $E_{\rm p} = \gamma_{\rm p} m_{\rm p} c^2$ and $\gamma_{\rm p}$ is the proton Lorentz factor \cite{chernoglazov_high-energy_2023} (for a qualitative discussion see also \cite{zhang_fast_2021, zhang_origin_2023}). We consider that a fraction $\eta_{\rm p}$ of the Poynting luminosity is transferred to non-thermal protons. Similarly to Eq.~\ref{eq:Lx}, the bolometric energy injection rate into relativistic protons can be written as 
    \begin{gather}
        L_{\rm p} = \eta_{\rm p} S_{\rm P} A = \frac{\eta_{\rm p}}{\eta_{\rm X}} L_{\rm X}.
        \label{eq:Lp}
    \end{gather}
We consider that protons are accelerated in a prior phase in the upstream region, before injected in the coronal environment. As a result, we assume that the differential spectrum of protons injected into the corona per unit time by the acceleration process is also described by a broken power law \cite{chernoglazov_high-energy_2023, zhang_fast_2021, zhang_origin_2023},
\begin{equation}
       \frac{{\rm d^2}N_{\rm p}}{{\rm d}\gamma_{\rm p} {\rm d}t}= \dot{N}_{\rm p,0} \left\{
        \begin{array}{cc}
           \gamma_{\rm p, \rm br}^{-s_{\rm p} +1} \gamma_{\rm p}^{-1},  &  1\leq \gamma_{\rm p} \leq \gamma_{\rm p, \rm br} \\
            \gamma_{\rm p}^{-s_{\rm p}}, & \gamma_{\rm p, \rm br} <  \gamma_{\rm p} \leq \gamma_{\rm \max}.
        \end{array}
        \right. \label{eq:dNdEdt}
    \end{equation}
where $\dot{N}_{\rm p,0}$ is a normalization factor, $\gamma_{\rm p, \rm br}$ is the break Lorentz factor, and $s_{\rm p}$ is the post-break power-law slope, which is taken to be 2 or 3 (the choice and effect of these values will be discussed in Section \ref{sec:res}). The post-break slope can be affected by the strength of the guide field, see e.g. Ref.~\cite{2017ApJ...843L..27W}. The break Lorentz factor $\gamma_{\rm p, \rm br}$ is dictated by energy conservation arguments, namely the injected luminosity in the accelerated protons, $L_{\rm p}$ is at most comparable to the Poynting flux, $S_{\rm P}$, times the area of the current sheet. Assuming that all protons from the upstream plasma are accelerated by the reconnection process, it follows that $\gamma_{\rm p, \rm br} \simeq \sigma_{\rm p}$ \cite{com2024}, where the proton magnetization value is defined as 
\begin{equation}
\sigma_{\rm p} = \frac{B^2}{4 \pi n_{\rm p} m_{\rm p} c^2} = \sigma_{\rm e} \frac{n_{\rm \pm}}{n_{\rm p}} \frac{m_{\rm e}}{m_{\rm p}},
\label{eq:sigma_p}
\end{equation}
with $n_{\rm p}$ being the proton number density. The maximum energy of the proton distribution is set either by the size of the corona (Hillas confinement criterion, $E_{\rm p , \max} = e B R$ \cite{Hillas_1984}), or it is radiation-limited (i.e., the energy loss rate balances the acceleration rate).  
Since, in the cases we examine, the radiation-limited Lorentz factor is the lowest one, $E_{\rm p, \max}$ is determined by the balance between proton cooling and acceleration. 
    
Using Eq.~\ref{eq:dNdEdt} we may express the differential power of protons injected into the corona as
\begin{equation}
\label{eq:Lp-diff}
L_{\rm p} (E_{\rm p}) \equiv E_{\rm p} \frac{{\rm d}^2N_{\rm p}}{{\rm d}t {\rm d}E_{\rm p}} = L_0 \left \{ 
    \begin{array}{cc}
           1,  &  m_{\rm p} c^2 \leq E_{\rm p} \leq E_{\rm p, \rm br} \\
            \left(\frac{E_{\rm p}}{E_{\rm p, \rm br}}\right)^{-s_{\rm p}+1}, & E_{\rm p, \rm br} <  E_{\rm p} \le E_{\rm p,\max}.
        \end{array}
                \right.
\end{equation}
where $E_{\rm p, \rm br}= \sigma_{\rm p} m_{\rm p} c^2 \gg E_{\rm p, \min}$, and $L_0$ is a normalization that relates to $L_{\rm p}$ as
\begin{equation}
\label{eq:L0}
L_0 \approx  L_{\rm p} \left \{ 
\begin{array}{cc}
E_{\rm p, \rm br}^{-1}\left(1+\ln(\frac{E_{\rm p,\max}}{E_{\rm p, \rm br}})\right)^{-1}, & s_{\rm p}=2 \\ 
(2 E_{\rm p, \rm br})^{-1}, & s_{\rm p}=3.
\end{array}
                \right.
\end{equation}

\section{Analytical Estimates}\label{sec:analytic}

In this section we derive analytical expressions for the neutrino energy spectrum, and the total secondary pair density produced in the coronal region due to proton-photon and photon-photon interactions, as a function of the black hole mass and the coronal X-ray luminosity. We also examine for what parameter values the proton power transferred to secondary populations is saturated (calorimetric limit). Our analytical expressions, which are derived under several simplifying assumptions, will be compared against numerical results in Section \ref{sec:res}.

\subsection{Neutrino luminosity and peak energy}\label{sec:analytic-neutrino}

High-energy neutrinos are produced through interactions of protons with X-ray photons from the corona. We can estimate the neutrino luminosity as
\begin{gather}
E_{\rm \nu} L_{\rm \nu+\bar{\nu}}(E_{\rm \nu}) \approx \frac{3}{8} f_{\rm p\gamma}(E_{\rm p}) E_{\rm p} L_{\rm p}(E_{\rm p}) \large |_{\rm E_{\rm p} \approx 20 E_{\rm \nu}},
\label{eq:Lnu}
\end{gather}
where $f_{\rm p\gamma}$ is the photomeson production efficiency for a proton of energy $E_{\rm p}$, and is defined as
    \begin{gather}
        f_{\rm p\gamma}(E_{\rm p})=\min \left \{ 1, t_{\rm p\gamma}^{-1}(E_{\rm p}) \frac{R}{c} \right \}. \label{eq:f_pg}
    \end{gather}
The corona is said to be optically thick to photomeson production processes when the first term in the curly bracket dominates, and optically thin otherwise. For the $ p\gamma$ energy-loss timescale $t_{\rm p\gamma}$, we can use Eq.~9 from Ref.~\cite{2024ApJ...961L..14F} and write the second term in the curly brackets as
\begin{gather}
t^{-1}_{\rm p\gamma}(E_{\rm p})\frac{R}{c} = 4\hat{\sigma}_{\rm p\gamma} \frac{ L_{\rm X}}{m_{\rm e} c^3} \frac{1}{9\ln(10) \tilde{R} R_{ \rm g} \epsilon_{\rm r}} \frac{\min( E_{\rm p}, E_{\rm p, *})}{m_{\rm p} c^2} \simeq 0.3 \frac{\lambda_{\rm X, Edd, -2}} {\tilde{R}}\frac{ \min( E_{\rm p}, E_{\rm p, *})}{25\, \rm TeV}
\label{eq:tpg}
\end{gather}
where $\epsilon_{\rm r} = 390$ is the energy threshold of the interaction (in units of $m_{\rm e} c^2$), $\hat{\sigma}_{\rm p\gamma} \approx 70~\mu b$ is the effective cross section \cite{2003ApJ...586...79A}, and $E_{\rm p, *}$ is the characteristic energy of protons interacting at threshold with the lowest energy coronal photons,
\begin{gather}
E_{\rm p,*} = \frac{\epsilon_{\rm r} m_{\rm p} m_{\rm e} c^4}{2 \epsilon_{\rm X, \min}} \simeq 930~\rm TeV \left(\frac{\epsilon_{\rm X, \min}}{0.1~\rm keV} \right)^{-1}.
\label{eq:Epstar}
\end{gather}
When $E_{\rm p} > E_{\rm p, *}$ the photomeson production efficiency becomes approximately independent of the proton energy. Substitution of expressions \ref{eq:f_pg} and \ref{eq:tpg} into Eq.~\ref{eq:Lnu} leads to
\begin{gather}
E_{\rm \nu} L_{\rm \nu+\bar{\nu}}(E_{\rm \nu}) \approx \frac{3}{8}   \min  \left \{1, 0.3 \frac{\lambda_{\rm X, Edd, -2}} {\tilde{R}}\frac{\min \left(E_{\rm p}, E_{\rm p,*}\right) }{25\, \rm TeV} \right\}E_{\rm p} L_{\rm p}(E_{\rm p}) \large |_{\rm E_{\rm p} \approx 20 E_{\rm \nu}}
\end{gather}
or, after using Eqs.~\ref{eq:Lp-diff} and \ref{eq:L0} and $s_{\rm p}=3$, 
\begin{equation}
E_{\rm \nu} L_{\rm \nu+\bar{\nu}}(E_{\rm \nu}) \approx \frac{3\eta_{\rm p}}{16\eta_{\rm X}} L_{\rm X}  \min  \left \{1, 1.2 \frac{\lambda_{\rm X, Edd, -2}} {\tilde{R}}\frac{\min \left(E_{\rm \nu}, E_{\rm \nu,*}\right) }{5\, \rm TeV} \right\} 
\left \{ 
\begin{array}{cc}
\dfrac{E_{\rm \nu}}{E_{\rm \nu, \rm br}} &, E_{\rm \nu} \le E_{\rm \nu, \rm br} \\
\left(\dfrac{E_{\rm \nu}}{E_{\rm \nu, \rm br}}\right)^{-1} &, E_{\rm \nu} > E_{\rm \nu, \rm br} \\
\end{array}
\right. \label{eq:nuLnu}
\end{equation}
where $ E_{\rm p} \approx 20  E_{\rm \nu}$,  $ E_{\rm p, *} \approx 20  E_{\rm \nu, *}$ and $E_{\rm \nu, \rm br} \approx 5 \sigma_{\rm p, 5}$~TeV. When evaluated at $ E_{\rm \nu}=E_{\rm \nu, \rm br}$, Eq.~\ref{eq:nuLnu} provides an approximation for the bolometric neutrino luminosity, which is derived in Appendix \ref{app:Lnu-bolo}). For coronae optically thin to photomeson production, we expect $E_{\rm \nu,\rm br} L_{\rm \nu}(E_{\rm \nu, \rm br}) / L_{\rm X} \propto \lambda_{\rm X, Edd}$, or to be constant otherwise. 

\subsection{Calorimetric limit for $p\gamma$ interactions}\label{sec:calorimetric}
When the efficiency of $ p\gamma$ interactions becomes equal to unity, then all the available energy of the parent proton population is transferred to secondary particles, such as leptons and neutrinos. Any further increase in the number density of target photons, for fixed proton luminosity, does not lead to an increase of the secondary injection rate. This saturation indicates the \textit{calorimetric limit} of protons in the corona (equivalently the corona becomes optically thick to $p\gamma$). 

Starting from the condition $f_{\rm p\gamma}(E)|_{E=\min (E_{\rm p, br}, E_{\rm p, *})}=1$ and using Eq.~\ref{eq:f_pg} we find that the calorimetric limit is achieved when the Eddington ratio of the source is
 \begin{gather}
      \lambda_{\rm Edd, crit} \simeq 0.03 \tilde{R} \left(\frac{\min(E_{\rm p, \rm br}, E_{\rm p,*})}{25~\rm TeV} \right)^{-1}.
     \label{eq:ledd_cal_lim} 
\end{gather}
This critical value of the Eddington ratio scales as $\propto \sigma_{\rm p}^{-1}$, as long as $E_{\rm p, br}< E_{\rm p, *}$, since $E_{\rm p, br} \propto \sigma_{\rm p}$. Alternatively, we can define a critical neutrino energy, $E_{\nu, \rm crit}$,  that indicates if the proton population in the corona is calorimetric to $p\gamma$ interactions for a given value of $\lambda_{\rm X, Edd}$. Using the relation $E_{p} \approx 20 E_{\nu}$ and re-arranging Eq.~\ref{eq:ledd_cal_lim} we find
\begin{equation}
    E_{\rm \nu, crit} \simeq 4\;\mathrm{TeV}\;\frac{\tilde{R}}{\lambda_{\rm X, Edd,-2}}.
\end{equation}
where we assumed that $ E_{\rm \nu, br} \leq E_{\rm \nu, *}$ (this holds for $\sigma_p \leq 10^6$).

If $E_{\rm \nu}^{\mathrm{crit}}\lesssim E_{\rm \nu,\rm br}$, the neutrino production at its peak is already calorimetric, meaning that the majority of the proton energy distribution is transferred to secondaries.

\subsection{Secondary pair number density}\label{sec:analytic-pairs}
In order to estimate the total density of secondary pairs created in the corona environment indirectly through hadronic interactions, we take into account three pair creation channels, as outlined below\footnote{The symbol $\gamma$ without subscript is used to indicate the target photons for each interaction.}:

\begin{tikzpicture}
    \centering
    \node (pg) at (-2, 2.){
    (i) $
    p^+ + \gamma \rightarrow 
    \left \{ \hspace{-1.5cm}
    \begin{array}{c}
         p^{+}+ \pi^{0} \rightarrow \gamma_{\rm TeV} + \gamma_{\rm TeV} \xrightarrow{\gamma-\gamma \rm ~annihilation} e^{+}+ e^{-} \\
         \rm or\\
         n^0 + \pi^{+} \rightarrow 
         \mu^{+} + 
         \bar \nu_{\rm \mu} \\
          \rm or\\
         p^{+} + \pi^{-} +\pi^{+} \rightarrow 
         \mu^{+}+ 
         \bar \nu_{\rm \mu} + \mu^{-}+ 
          \nu_{\rm \mu} \\ {} \\
        \hspace{7.cm} \mu^{+} (\mu^{-}) \rightarrow e^{+} (e^{-}) + \bar \nu_{e} (\nu_{e}) + \nu_{\mu} (\bar \nu_{\mu})
         % \mu^{+} (\mu^{-}) \rightarrow e^{+} (e^{-}) + \bar \nu_{e} (\nu_{e})
         %\nu_{\rm \mu} + \nu_{\rm e} (\bar \nu_e) + e^{+} (e^{-})
    \end{array}
    \right.
    $
    };

    \node[below=0.5 cm of pg] (BH) {
    (ii) $
    p^+ +\gamma \rightarrow p^+ + e^{+}+e^{-}
    $
    };

    \node[below=0.5 cm of BH] (gg){
    (iii) $
    \gamma_{\rm MeV} + \gamma \rightarrow e^{+} +e^{-}
    $
    };

    \draw[->,>=stealth', auto, left=20, node distance=3cm, thick,] (-1.5, 1.2) to (-1.5, .5) to (-1., .5);
    
    \draw[->,>=stealth', auto, left=20, node distance=3cm, thick,] (1., 0.2) to (1., 0.1) to (1.9, 0.1) to (1.9, 0.2) to (1.9, 0.1) to (1.45, 0.1) to (1.45, -1.5) to (-3., -1.5) to (-3., -1.8);

    \draw[->,>=stealth', auto, left=20, node distance=3cm, thick,] (-1., -1.) to (-1., -1.1) to (-.0, -1.1) to (-.0, -1.) to (-.0, -1.1) to (-.5, -1.1) to (-.5, -1.1) to (-.5, -1.5) to (-3., -1.5) to (-3., -1.8);

    \node (cool) at (-2.2, -1.3){$ \rm cool$};
    
\end{tikzpicture}

The first ($ p\gamma$) channel injects pairs in the corona indirectly via the attenuation of TeV photons produced in neutral pion decays, generated by relativistic protons, and directly via the decay of charged pions. Additionally, Bethe-Heitler proton-photon interactions (second channel) act as a direct pair production channel. Lastly, the third ($ \gamma \gamma$ annihilation)  channel  injects pairs through the attenuation of MeV photons resulting from the electromagnetic cascade (namely, photons emitted by the relativistic pairs injected in channels (i) and (ii)). Because of radiative losses the injected relativistic pairs cool down reaching Lorentz factors of $\gamma_{\rm e}\sim 1$, forming the pool of non-relativistic pairs. 

Simple physical arguments show that the $ \gamma \gamma$ annihilation channel vastly dominates over the others in terms of pair production. $ p\gamma$ injects an energy content in TeV photons which is nearly immediately reprocessed into a comparable energy content in MeV photons. However, for the same energy content, the number density of TeV photons injected is much smaller than the number density of MeV photons, by about 6 orders of magnitude. Therefore, the amount of pairs that can be injected by $ p\gamma$ pair production directly is roughly suppressed by 6 orders of magnitude compared to the $\gamma\gamma$ injection channel. Bethe-Heitler injection is also subdominant compared to the $ \gamma \gamma$ annihilation channel, because the energy injected directly into pairs by the former channel is less than the energy ending up to MeV photons. In Appendix \ref{App:channels}, we explicitly evaluate the amount of pairs injected by $ p\gamma$ and Bethe-Heitler, confirming these order-of-magnitude estimates. In the following, we present in detail the predictions for the $\gamma\gamma$ annihilation channel.

\subsubsection{The $ \gamma \gamma$ annihilation channel}
The annihilation of photons with energies $E_{\rm \gamma} \sim 1-10$~MeV (henceforth, soft $ \gamma$ rays) will lead to the creation of pairs. Assuming that each particle takes half of the energy of the $ \gamma$-ray photon, $E_{\rm \pm} \approx E_{\rm \gamma}/2$, and that all of the soft $ \gamma$-ray luminosity is transferred to pairs\footnote{This implies that the corona is optically thick to soft $ \gamma$-rays. In Appendix \ref{App:gg} we present numerical results for the $ \gamma \gamma$ opacity of the corona supporting our assumption.},  $L_{\rm MeV} \approx L^{\gamma \gamma}_{\rm \pm}$, we can calculate the injection rate of relativistic pairs as
    \begin{gather}
            \dot n_{\rm \pm}= \frac{L^{\gamma \gamma}_{\rm \pm}}{E_\pm V}, \label{eq:n_pm_dot}
    \end{gather}
where $V= 4\pi R^3/3$ is the corona volume. Because the cooling timescale of the $ \gamma \gamma$ injected pairs is short compared to the particle escape timescale from the corona\footnote{In most cases synchrotron cooling dominates the radiative losses of pairs, and $ ct_{\rm syn}/R = 6 \pi m_{\rm e} c^2/(\sigma_{\rm T} B^2 R) (\gamma_{\rm e}/10)^{-1} \ll 1$. This is also validated by our numerical results in section \ref{sec:res}.}, which is taken to be $ R/c$, Eq.~\ref{eq:n_pm_dot} can also be used for the injection rate of non-relativistic pairs.
Then, the number density of non-relativistic pairs can be estimated as 
    \begin{gather}
            n_{\rm \pm}= \frac{R}{c} \dot n_{\rm \pm} = \frac{3L_{\rm MeV}}{2 \pi R^2 c E_\gamma}.\label{eq:n_pm}
        \end{gather}

The next step is the estimation of the MeV photon luminosity. The process that dominates the production of soft $ \gamma$-ray photons depends on the system parameters, like $ \sigma_{\rm p}$ (this will become clearer by our numerical results presented in the next section). One possibility is that the synchrotron emission of Bethe-Heitler pairs dominates the MeV range (see for example figure~\ref{fig:components}). Using approximate expressions for the relevant effective cross sections, it can be shown that $ L_{\rm \nu+\bar \nu}/L_{\rm BH, syn} \approx 33 \, (n_{\rm X, \rm p\gamma}/n_{\rm X, \rm BH})$, where $ n_{\rm X, \rm p\gamma}$ and $ n_{\rm X, \rm BH}$ are the target photon number densities for photomeson and Bethe-Heitler production processes respectively (for more details see Ref.~\cite{karavola_BH_2024}). When $ n_{\rm X}(E) \propto E^{-2}$, as assumed here, then $ L_{\rm \nu+\bar \nu}/L_{\rm BH, syn} \approx 33 \,  \left(E_{\rm X, \rm BH}/E_{\rm X, \rm p\gamma} \right) \approx 0.23$ (i.e. of order unity), with $ E_{\rm X, \rm BH}=2 m_{\rm e} c^2/\gamma_{\rm p}$ and $ E_{\rm X, \rm p\gamma}=m_{\rm \pi}c^2/\gamma_{\rm p} \simeq 290 m_{\rm e} c^2/\gamma_{\rm p}$ being the target photon energies for on-threshold Bethe-Heitler and photomeson interactions respectively. Similar estimates can be made if the emission is dominated by the radiation of secondary pairs from photomeson interactions. Therefore, regardless of the origin of soft $ \gamma$-ray photons, we can use the approximate relation

\begin{equation}
L_{\rm MeV} \approx L_{\rm \nu+\bar \nu} \approx \frac{3}{8} f_{\rm p\gamma}(E_{\rm p, \rm br}) E_{\rm p, \rm br} L_{\rm p}(E_{\rm p, \rm br}).
\label{eq:LMeV}
\end{equation} 

Using Eqs.~\ref{eq:n_pm} and \ref{eq:LMeV} the number density of the pairs produced by the interaction of photons with energy $E_{\rm \gamma}$ with the coronal ones can be estimated as: 
\begin{gather}
    n_{\rm \pm}^{\gamma\gamma}=\frac{9\eta_{\rm p}}{32 \pi \eta_{\rm X} }  \frac{f_{\rm p\gamma} L_{\rm X}}{E_\gamma  R^2c} =  5.4 \cdot 10^{12} {\rm ~cm^{-3}}  \frac{\lambda_{\rm X, Edd,-2}^2}{\tilde{R}^2 L_{\rm X,43}} \left(\frac{10~{\rm MeV}}{E_\gamma}\right)  \min \left \{ 1, 0.3 \frac{\lambda_{\rm X, Edd,-2}}{\tilde{R}} \frac{\min(E_{\rm p, \rm br}, E_{\rm p}^*)}{25~{\rm TeV}} \right \}. \label{n_ee_gg-new}
\end{gather}
where we used 10~MeV as the nominal value for the $ \gamma$-ray photon energy, where the corona becomes optically thick to $ \gamma \gamma$ annihilation (for more details, see Appendix \ref{App:gg}). 

The total pair density can be estimated by the sum of the terms $ p\gamma$ (Eq.~\ref{n_ee_pg}),  Bethe-Heitler (Eq.~\ref{n_ee_BH-new}) and $ \gamma \gamma$ annihilation (Eq.~\ref{n_ee_gg-new}) terms. However, inspection of the three aforementioned expressions shows that the attenuation of soft $ \gamma$-ray photons ($1-10$ MeV) is the most important contributor to the density of secondary pairs, in agreement with the qualitative arguments presented above.

\section{Numerical results} \label{sec:res}

In this section we present the numerical results of our parametric study, which was performed using the leptohadronic numerical code \code \ \cite{Dimitrakoudis, mastichiadis_spectral_2005}. The code solves a system of coupled partial differential equations for all stable particle species inside a spherical, magnetized region, namely electrons and positrons (pairs), protons, neutrons\footnote{Neutrons have a decay time of $\sim$15 minutes, in their own rest frame. However, they are considered stable particles, in our calculation, due to the fact that in the lab rest frame they have a decay timescale much longer than the escape time, $R/c$.}, photons, and neutrinos. All parameter values are listed in Table \ref{tab:params}.

\begin{table*}[h!]
    \centering
\caption{Model parameters. The top three are the key parameters that we vary.}
 \label{tab:params}
    \begin{tabular}{l c c}
    \hline
    Parameter & Symbol (unit) & Value range  \\ 
    \hline
    Black hole mass & $M_{\rm bh} \, (M_{\rm \odot})$ & $10^5-10^9$ \\ 
    Broadband (0.1-100 keV) X-ray luminosity & $L_{\rm X}$ (erg s$^{-1}$) &  $10^{40}-10^{44}$ \\ 
    Proton plasma magnetization & $ \sigma_{\rm p}$ & $10^3-10^7$ \\ 
    \hline
    Proton post-break slope & $s_{\rm p}$ & 2 or 3 \\ 
    Maximum proton Lorentz factor & $\gamma_{\rm p, \max}$ & $10^8$ \\ 
    Size of reconnection layer & $R$ ($R_{\rm g}$)  &  3 \\
    Reconnection rate & $\beta_{\rm rec}$ & 0.1 \\
    X-ray photon index & $s_{\rm X}$ & 2 \\ 
    X-ray to magnetic energy density ratio & $\eta_{\rm X}$ & 0.5 \\ 
    Non-thermal proton to Poynting luminosity ratio & $\eta_{\rm p}$ & 1/3 \\ 
    \hline 
    \end{tabular}
\end{table*}

\begin{figure}
    \centering
    \includegraphics[width=0.49\linewidth]{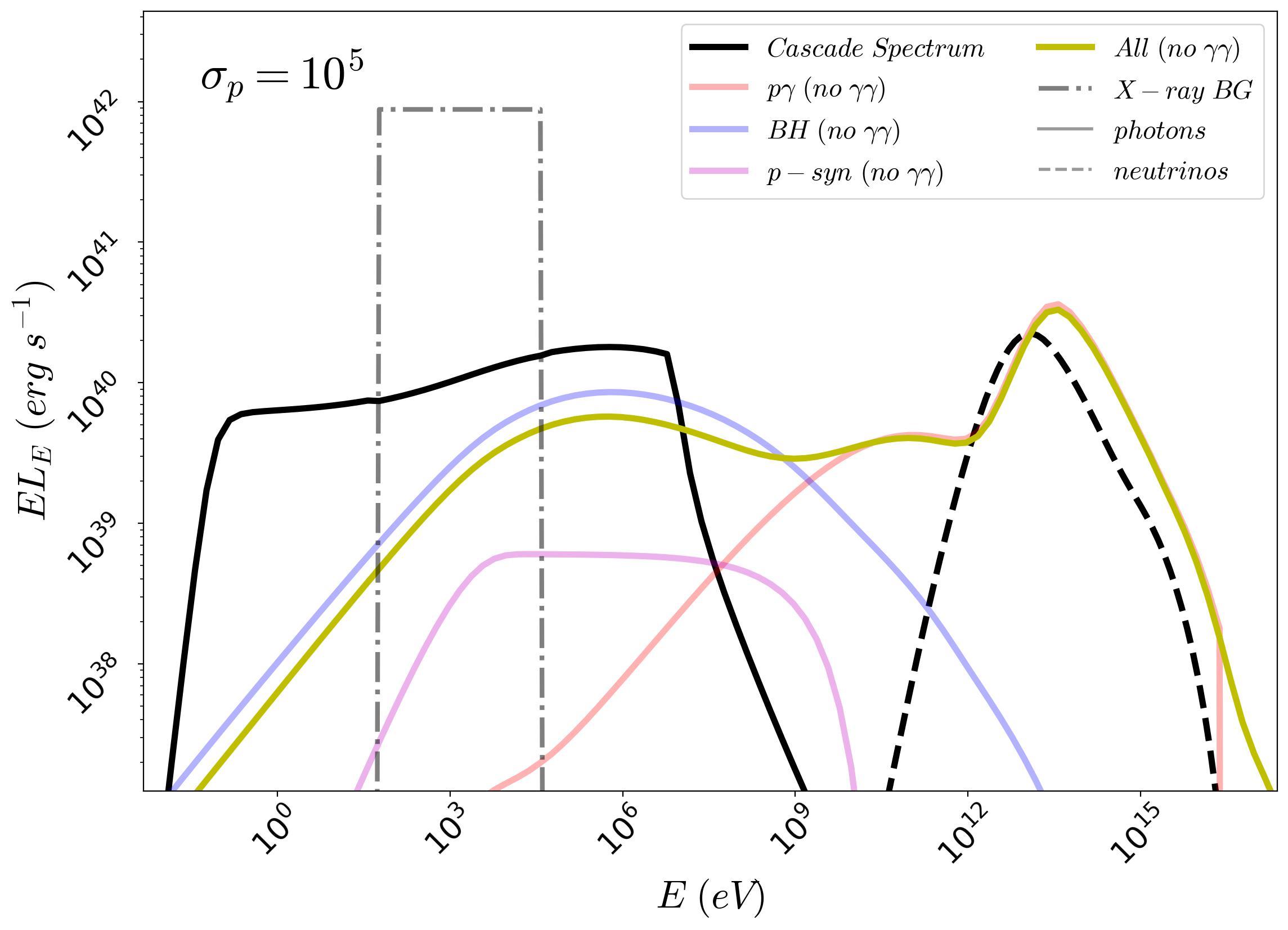}
    \includegraphics[width=0.49\linewidth]{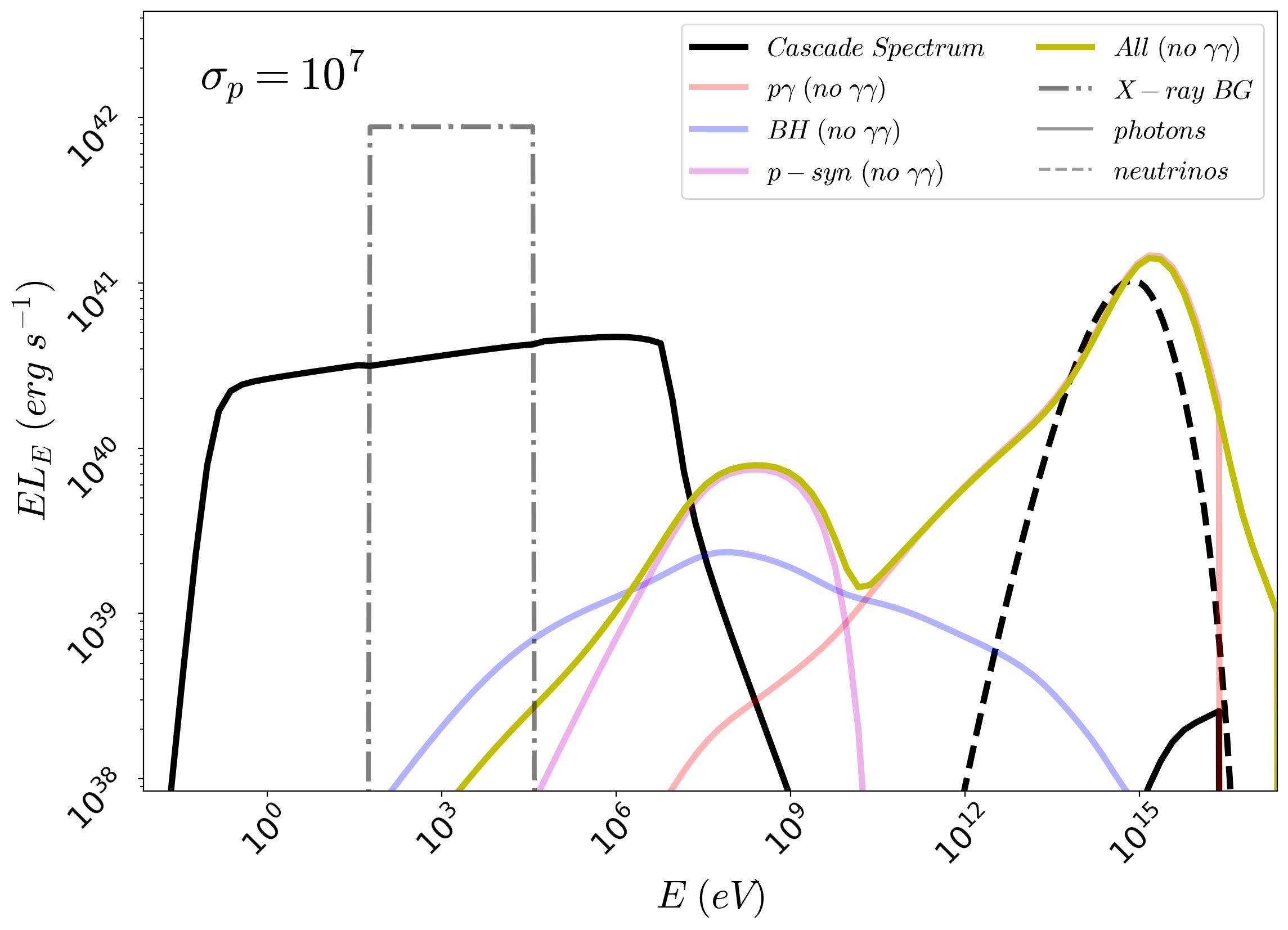}
    \caption{Decomposition of the spectral energy distribution (SED) of the hadronic cascade emission (black solid line) for $\sigma_{\rm p}=10^5$ (left panel) and $\sigma_{\rm p}=10^7$ (right panel). The black dashed line represents the all-flavor neutrino energy distribution. The dash-dotted grey line indicates the corona X-ray spectrum. Other parameter used here are: $s_{\rm p}=2$, $L_{\rm X}=10^{43}$~erg s$^{-1}$ and $M_{\rm bh}=10^7 M_\odot$ ($\lambda_{\rm X, Edd}=10^{-2}$). 
    }
    \label{fig:components}
\end{figure}

In figure \ref{fig:components} we show indicative electromagnetic (solid lines) and all-flavor neutrino (dashed lines) spectra for the AGN coronal region as described by our model for two values of $ \sigma_{\rm p}$. Black solid line shows the photon spectrum escaping the corona with all leptonic and hadronic processes contributing to it, while yellow solid line represents the same spectrum without accounting for $\gamma \gamma$ annihilation. Moreover, each of the other colored lines represents one individual hadronic spectral component: the orange line refers to the emission from secondary pairs produced in photomeson interactions, the blue one indicates the Bethe-Heitler pair emission component, and the magenta one corresponds to the proton synchrotron spectrum. 

\begin{figure}
\centering 
\includegraphics[width=0.65\textwidth]{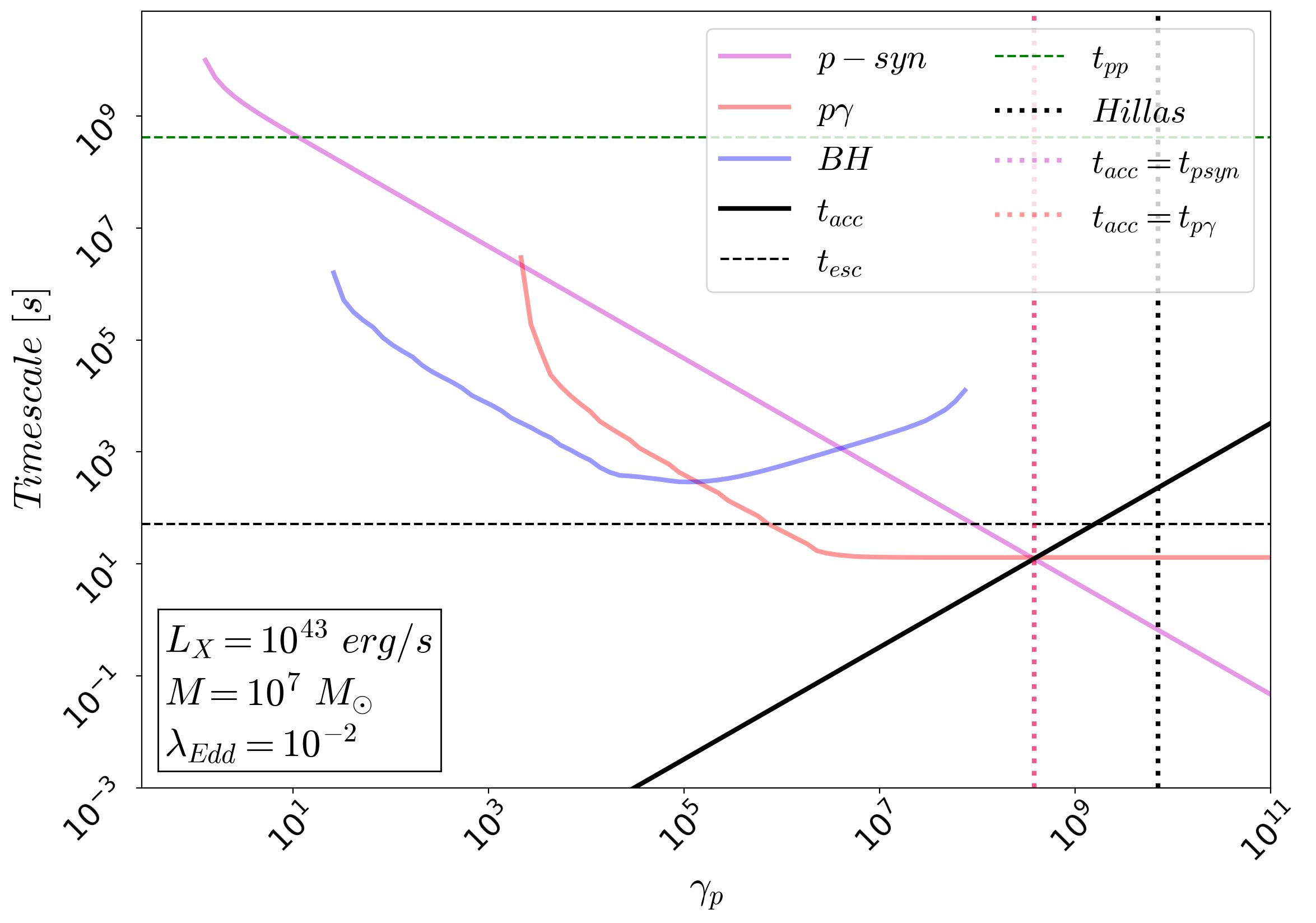}
\caption{
Characteristic timescales as a function of the proton Lorentz factor for the parameters of figure \ref{fig:components}. Solid lines represent the timescales for the acceleration process (black),  proton synchrotron cooling (magenta), photomeson production (orange), and Bethe-Heitler pair production (blue). The X-ray photons of the corona are considered as targets for the two latter processes. The dotted black line represents the Hillas-limited Lorentz factor, while magenta and orange dotted lines represent the proton Lorentz factor at which the acceleration timescale is equal to the synchrotron and photomeson ones, respectively.}
\label{fig:timescales}
\end{figure}

We observe that the cascade spectra escaping the corona (solid black lines) in both panels are similar. The aforementioned characteristic is because the escaping photon spectrum is shaped by photon-photon pair creation, which has no direct dependence on $ \sigma_{\rm p}$. In particular, high-energy photons ($ E \gtrsim 10$~MeV) interact with lower energy ones creating relativistic electron-positron pairs that cool radiatively mainly due to synchrotron losses, resulting in an extended and almost flat spectrum up to 10 MeV (black line). However, the contribution of the individual hadronic components to the total spectrum is vastly different in the two cases. For $ \sigma_{\rm p} =10^5$, 
protons that carry most of the energy of the population, i.e., those with energy $ E_{\rm p, \rm br}=\sigma_{\rm p} m_{\rm p} c^2$, interact with X-ray photons of energy 0.1~keV close to the threshold energy for Bethe-Heitler pair production. For such interactions the proton loss timescale due to Bethe-Heitler pair production is the shortest, and also comparable to the photomeson loss timescale -- see figure~\ref{fig:timescales}. As a result, the energy transferred from protons with energy $E_{\rm p, \rm br}$ to secondary leptons is similar for the two interactions, hence their radiative output will be comparable -- see figure \ref{fig:components}. The Bethe-Heitler synchrotron spectrum has a well-defined peak at $\sim 1$~MeV and is produced by pairs injected with $ \gamma_{\rm e} \approx \sigma_{\rm p}$ (for more details about the Bethe-Heitler spectra, see \cite{karavola_BH_2024}). Meanwhile, pion decays produce on average more energetic pairs with $ \gamma_{\rm e} \approx 0.05 \, (m_{\rm p}/m_{\rm e}) \sigma_{\rm p}$ (see Eq.~\ref{eq:EE_pg} and \cite{Petropoulou_2015}) that emit photons of energy $\sim 10$~GeV. 

The proton synchrotron spectrum is subdominant, as the relevant loss timescale is longer than the other processes. The relative importance of proton synchrotron radiation, Bethe-Heitler pair production and photomeson production in the cooling of protons changes for higher proton break energies. When $\sigma_{\rm p}=10^7$, protons cool preferentially via the photomeson production process, which has the shortest timescale. Note that the photomeson timescale becomes almost independent of the proton energy for interactions away from the threshold value of the interaction (see also Eq.~\ref{eq:Epstar}), as opposed to the Bethe-Heitler timescale (its effective cross section increases logarithmically with  interaction energy \cite{mastichiadis_spectral_2005}). Moreover, the proton synchrotron timescale becomes comparable to the photomeson loss timescale. Therefore, for $ \sigma_{\rm p}=10^7$ we expect that the synchrotron spectra from protons and pairs from pion decays will have a comparable contribution to the total emission, while the Bethe-Heitler synchrotron component should be subdominant, as indeed shown in the right panel of figure~\ref{fig:components}. We note that in order to show the Bethe-Heitler synchrotron component separately, we re-run \code \ only with electron synchrotron and Bethe-Heitler pair production on. As a result, protons do not lose part of their energy due to proton synchrotron or photopion production (as they normally do -- see black and yellow curves). As a result, there is more energy available for the Bethe-Heitler produced pairs, resulting in the blue curve being slightly higher than the yellow one (which includes all the interactions apart from $\gamma \gamma$ annihilation). 

\begin{figure}
\centering
        \includegraphics[width=0.45\textwidth]{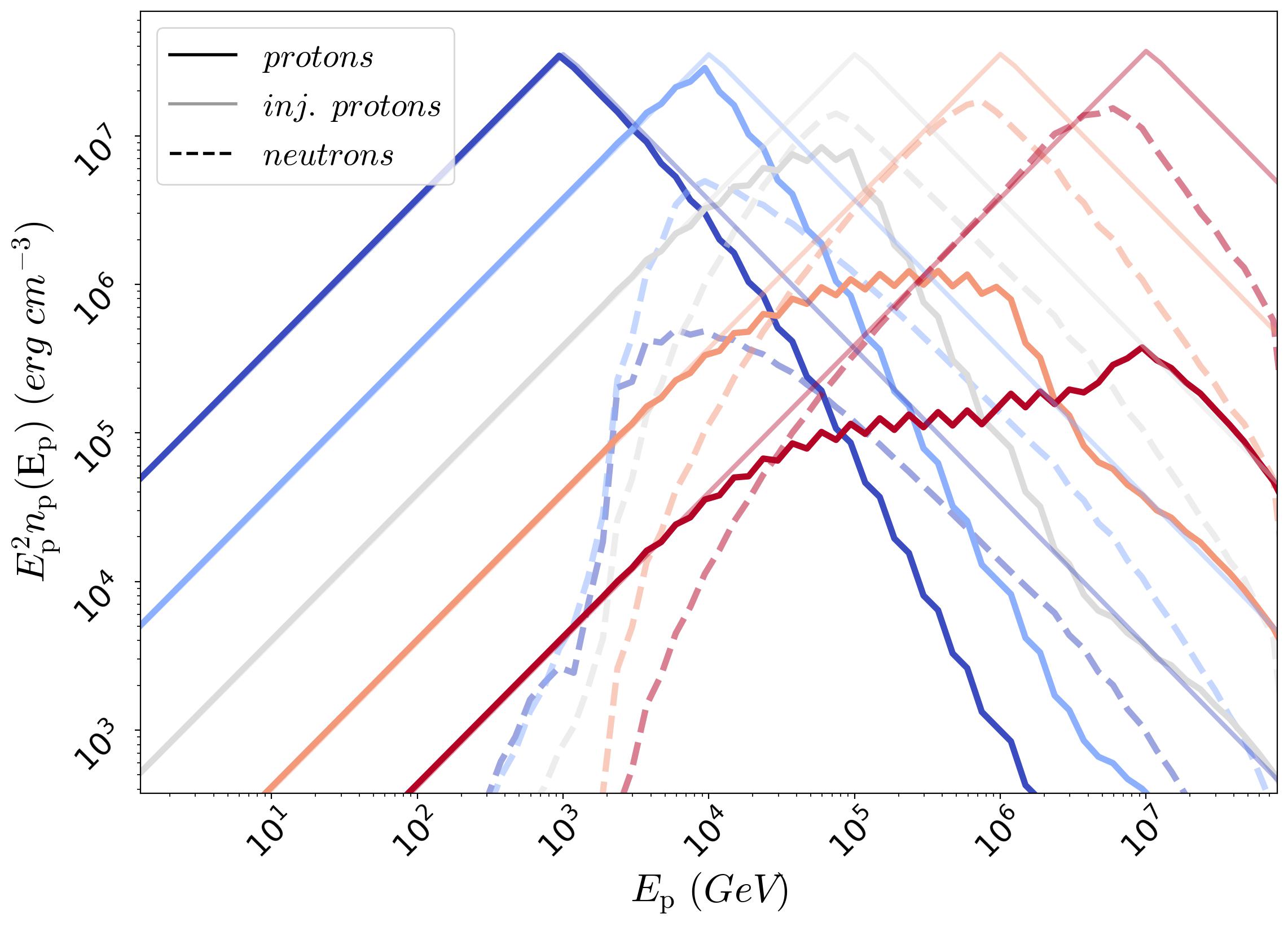}
        \includegraphics[width=0.495\textwidth]{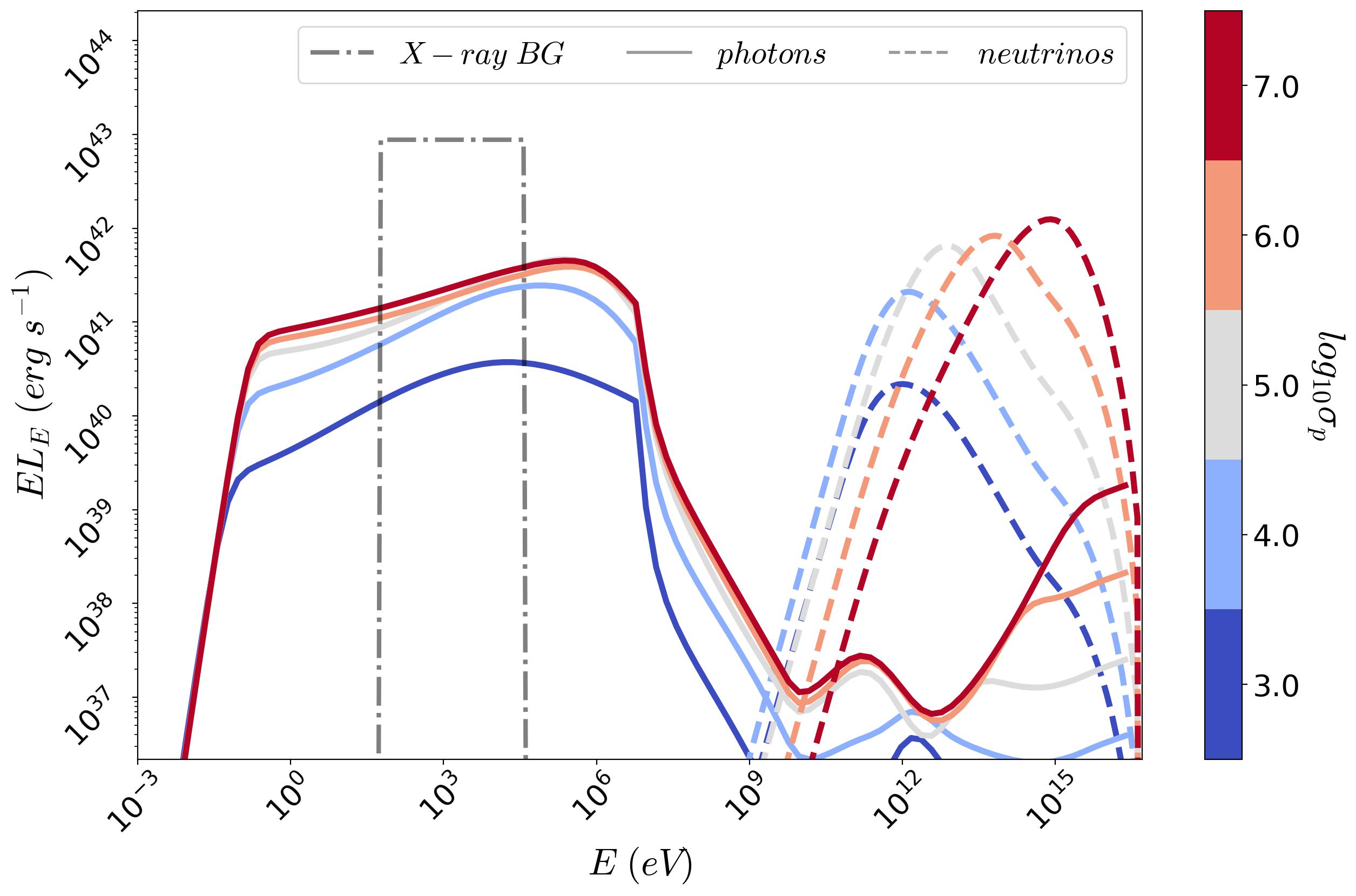}
        \captionof{figure}{\textit{Left panel:} Energy distribution of protons at injection (transparent solid lines) and at steady state (bold solid lines). Steady state neutron energy distributions are also shown (transparent dashed lines). The color bar refers to the proton magnetization value, which determines the peak of the injected proton energy distribution (see Eq.~\ref{eq:dNdEdt}). \textit{Right panel:} Spectral energy distribution of cascade photons (solid lines) and  neutrinos of all flavors (dashed lines) emitted from the coronal region.  Other parameters used: $ L_{\rm X}=10^{43}$~erg s$^{-1}$ and $ M_{\rm bh}=10^{7}M_{\rm \odot}$ ($\lambda_{\rm X, Edd}=10^{-2}$).}
        \label{fig:spectr_neutr_prot}
\end{figure}

Next we study the impact of $ \sigma_{\rm p}$ on the photon cascade and neutrino emission from the AGN coronal region. Figure \ref{fig:spectr_neutr_prot} shows our numerical results for $ L_{\rm X}=10^{43}$~erg s$^{-1}$, $M_{\rm bh}=10^{7}M_{\rm \odot}$, $s_{\rm p}=3$, and varying $ \sigma_{\rm p}$ in the range $10^3-10^7$. We do not explore even lower $ \sigma_{\rm p}$ values as these would lead to peak neutrino production at energies $\ll 1$~TeV, where observations are dominated by the neutrino atmospheric background. In the left panel we present the injected and steady-state proton distributions, as well as the steady-state neutron distribution in the corona. In the right panel we display the electromagnetic and neutrino spectra emitted from the coronal region. As $ \sigma_{\rm p}$ increases the cascade spectrum becomes more luminous, because the production rate of secondary pairs is higher; the energy lost by the proton population increases with $ \sigma_{\rm p}$ as shown in the left panel (see also figure~\ref{fig:timescales}). Moreover, the electromagnetic spectra show a sharp cutoff at about 10 MeV due to $ \gamma \gamma$ pair production on the highest energy X-ray photons (100~keV) for all $ \sigma_{\rm p}$ values. Similarly, the dip of the spectrum at about 10~GeV is caused by the attenuation of $ \gamma$-ray photons by the 0.1~keV photons of the corona that have larger number densities (for more details about the attenuation of energetic photons, see Appendix~\ref{App:gg}). Such attenuation leads to creation of secondary leptonic population which later on radiate through synchrotron, producing broad and almost flat (in $ E L_{\rm E}$ representation) electromagnetic spectra ($E \in [10^{-3}, 10^{7}]~\rm eV$, see figure~\ref{fig:components}). Given that in our model $\gamma \gamma$ pair creation of the aforementioned populations occurs for all $ \sigma_{\rm p}$ values, AGN coronae with different luminosity or black-hole masses will be characterized by similar cascade spectra as those shown in figure~\ref{fig:spectr_neutr_prot}.

The peak of the neutrino energy distribution, which is determined by protons with energy $ E_{\rm p, \rm br} = \sigma_{\rm p} m_{\rm p} c^2$, moves to higher energies as $ \sigma_{\rm p}$ increases, as shown on the right panel of figure \ref{fig:spectr_neutr_prot}. Moreover, the peak neutrino luminosity increases for higher $ \sigma_{\rm p}$ values (see Eq.~\ref{eq:nuLnu}), for the same reason the cascade emission becomes overall more luminous: for higher $ \sigma_{\rm p}$ values, more energy is carried by protons of the distribution that fulfill the energy threshold condition for pion production, resulting in a larger fraction of proton energy transferred to secondary particles (neutrinos, photons, and pairs). As the coronal proton population tends to become calorimetric to photomeson interactions, the cascade and neutrino luminosities tend to saturate to a constant value. Furthermore, as $ \sigma_{\rm p}$ increases, the overall shape of the neutrino spectrum changes, with the extent of the power law above the peak neutrino energy decreasing. The latter is due to the fact that as $ \sigma_{\rm p}$ increases the break energy of the proton energy distribution moves towards its upper energy limit, which is set by radiative losses. 

\begin{figure}
    \centering
     \includegraphics[width=0.9\textwidth]{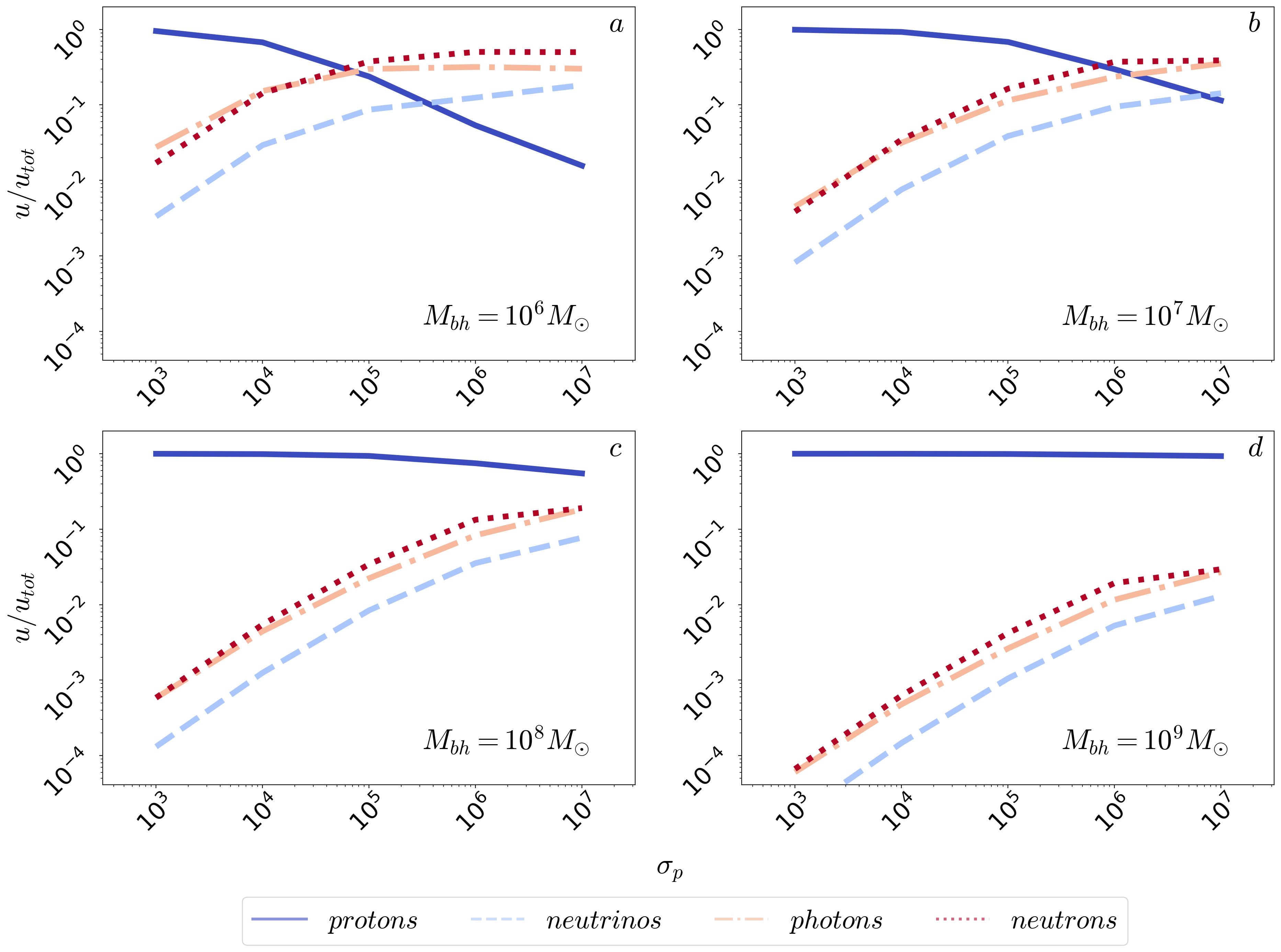}
        \captionof{figure}{Fraction of the proton (solid dark blue line), photon (dash-dotted orange line), neutrino (dashed light blue line), and neutron (dotted red line) bolometric energy density over the total energy density of all four species as a function of $ \sigma_{\rm p}$. The corona luminosity in all panels is $L_{\rm X} = 10^{43}$ erg s$^{-1}$ but the Eddington ratio changes from $10^{-1}$ in panel (a) to $10^{-4}$ in panel (d) in decrements of 10.}
    \label{fig:num_dens_frac}
\end{figure}

To demonstrate the impact of the Eddington ratio, $\lambda_{\rm X, Edd}$, we performed a series of numerical runs for an accreting system with $ L_{\rm X}=10^{43}$~erg s$^{-1}$ and $ M_{\rm bh} = 10^6 -10^9 M_{\rm \odot}$, and computed the energy densities of protons, and secondary particles at steady state in the corona for $ \sigma_{\rm p} = 10^3-10^7$. 
Our results are presented in figure \ref{fig:num_dens_frac}.
We find that as $\lambda_{\rm X, Edd}$ decreases (from panel a to panel d), the amount of energy transferred from the proton population to the secondary particles, namely photons, neutrons and neutrinos, also decreases for all values of $ \sigma_{\rm p}$. In the system with the lowest $\lambda_{\rm X, Edd}$ (panel d), protons do not lose a significant amount of their energy and thus, the system is optically thin to photohadronic interactions for all $ \sigma_{\rm p}$ values. On the contrary, in the system with the highest $\lambda_{\rm X, Edd}$ (panel a) protons transfer most of their energy to secondary populations, with most of it going to neutrons and photons. Furthermore, by comparing panels (a) and (b), we observe that the $ \sigma_{\rm p}$ value at which protons start to significantly lose energy increases almost linearly with $\lambda^{-1}_{\rm X, Edd}$ (see also Eq. \ref{eq:ledd_cal_lim}). For example, in panel (a) where $\lambda_{\rm X, Edd}= 10^{-1}$, the protons start to lose a noticeable fraction of their energy ($ \sim 0.5$ of their total energy) at $ \sigma_{\rm p} \approx 10^5$, while in panel (b) where $\lambda_{\rm X, Edd} = 10^{-2}$, the same happens for $ \sigma_{\rm p} \approx 10^6$, see Eq.~(\ref{eq:ledd_cal_lim}).
 
\begin{figure}
    \centering
    \includegraphics[width=0.99\textwidth]{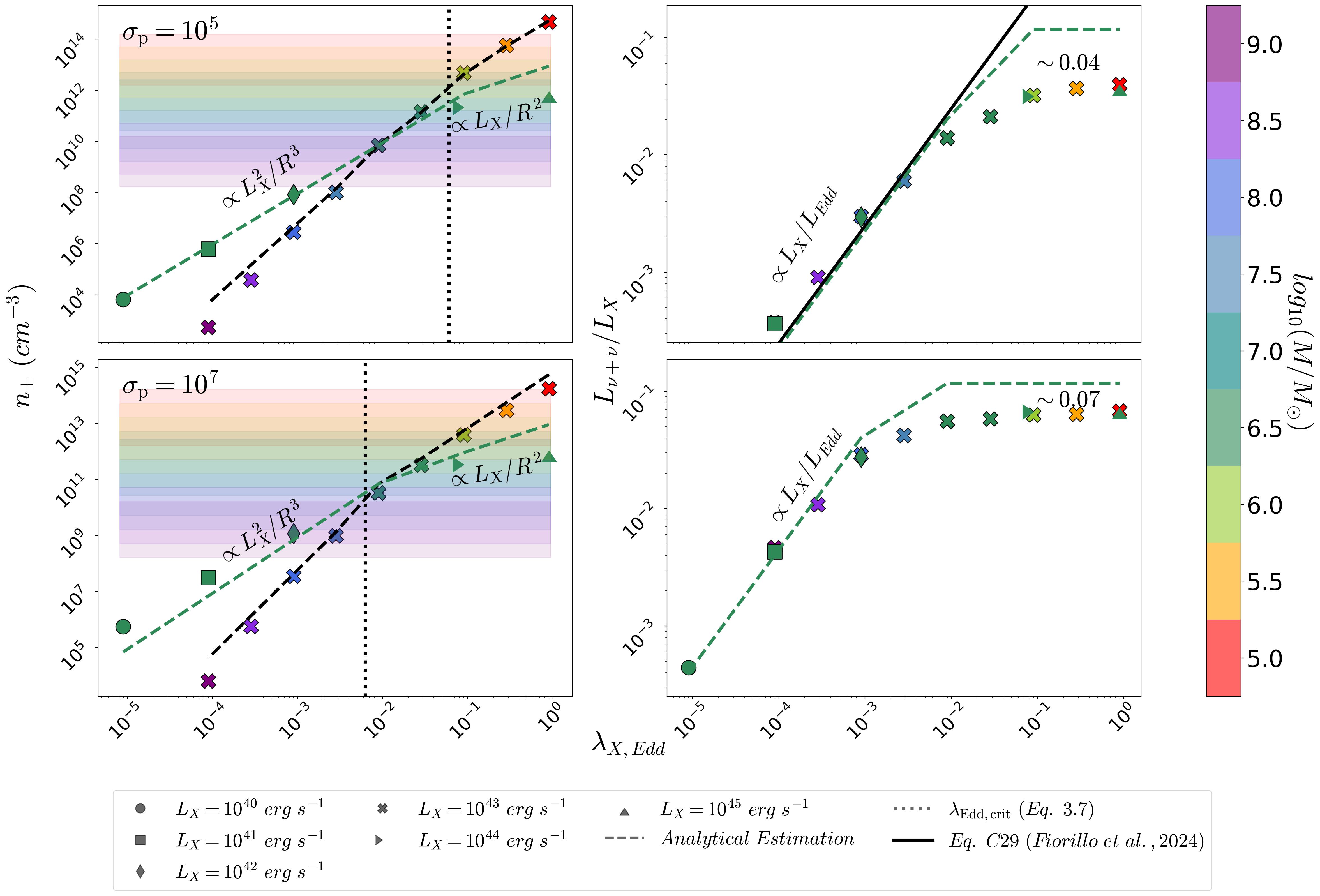}
        \caption{Non-relativistic secondary pair number density (left column) and all-flavor (bolometric) neutrino luminosity over corona X-ray luminosity (right column) as a function of the Eddington ratio (defined with respect to the broadband X-ray luminosity). Results for  $\sigma_{\rm p}=10^5$ and $10^7$ are shown in the upper and lower panels, respectively. }
    \label{fig:ne_max}
\end{figure}

We next examine whether proton-photon interactions and $ \gamma \gamma$ annihilation in the coronal region are able to create enough pairs to result in Thomson optical depths of 0.1 to 10. In order to do so, we show, in the left-hand side panels of figure \ref{fig:ne_max}, the secondary non-relativistic pair number density, $ n_{\rm \pm}$, computed numerically as a function of the Eddington ratio, $\lambda_{\rm X, Edd}$, for two values of $\sigma_{\rm p}$. Colored markers represent the numerical results of our parametric study for cases of (i) varying compact object mass, $ M_{\rm bh} =10^5-10^9 M_{\rm \odot}$ (mass range indicated by the colorbar) and (ii) varying coronal luminosity, $ L_{\rm X}=10^{40}-10^{44}~\rm erg\, s^{-1}$ (luminosity values plotted with different markers -- see legend of figure \ref{fig:ne_max}). Furthermore, the color-shaded bands represent the pair density values that correspond to $ \tau_{\rm T}=0.1-10$ for each black hole mass. Changes in either of these parameters affect the number density of X-ray photons in the corona, and in turn the opacities for proton-photon interactions and $\gamma \gamma$ pair production. When the secondary pair density is plotted against $\lambda_{\rm X, Edd}$ it can be described by a broken power-law function, with the break occurring when protons become calorimetric in the corona (see Eq.~\ref{eq:ledd_cal_lim} and vertical dotted line). For comparison purposes, we overplot the analytical expression for $ n_{\rm \pm}$ , as given by Eq.~\ref{n_ee_gg-new}, when $L_{\rm X}$ (dashed green lines) or $M_{\rm bh}$ (dashed black lines) is varying separately. We observe that our analytical approximation for the pair density in the coronal region, as given by Eq.~\ref{n_ee_gg-new}, is in good agreement with the numerical results, verifying our assumptions that the dominant source of pairs is $ \gamma \gamma$ annihilation of energetic photons produced indirectly via proton-photon interactions. The above result is valid both for the case where we vary the mass of the central black hole while we keep the broadband coronal luminosity fixed at $L_{\rm X}=10^{43} \rm erg~s^{-1}$ (color-changing crosses) and the scenario in which the coronal broadband luminosity varies while the black hole mass is fixed to $M_{\rm bh}=10^7 M_{\odot}$ (dark green shape-changing symbols). We also highlight that our estimation of the calorimetric limit describes well enough the change of slope of the dashed lines, marking the transition from an optically thin to an optically thick corona to $p\gamma$ interactions. In addition, we observe that the numerical $ n_{\rm \pm}$ values fall within their respective color-shaded bands for $\lambda_{\rm X, Edd}\gtrsim \lambda_{\rm Edd, crit}$, while for $\lambda_{\rm X, Edd} \lesssim \lambda_{\rm Edd, crit}$ the density of secondary pairs in the corona is not high enough as to result in optical depths of order unity. Therefore, the dominant source of pairs in the coronae of black holes accreting at low Eddington ratios cannot be of hadronic origin.

\begin{figure}
    \centering
    \includegraphics[width=0.8\linewidth]{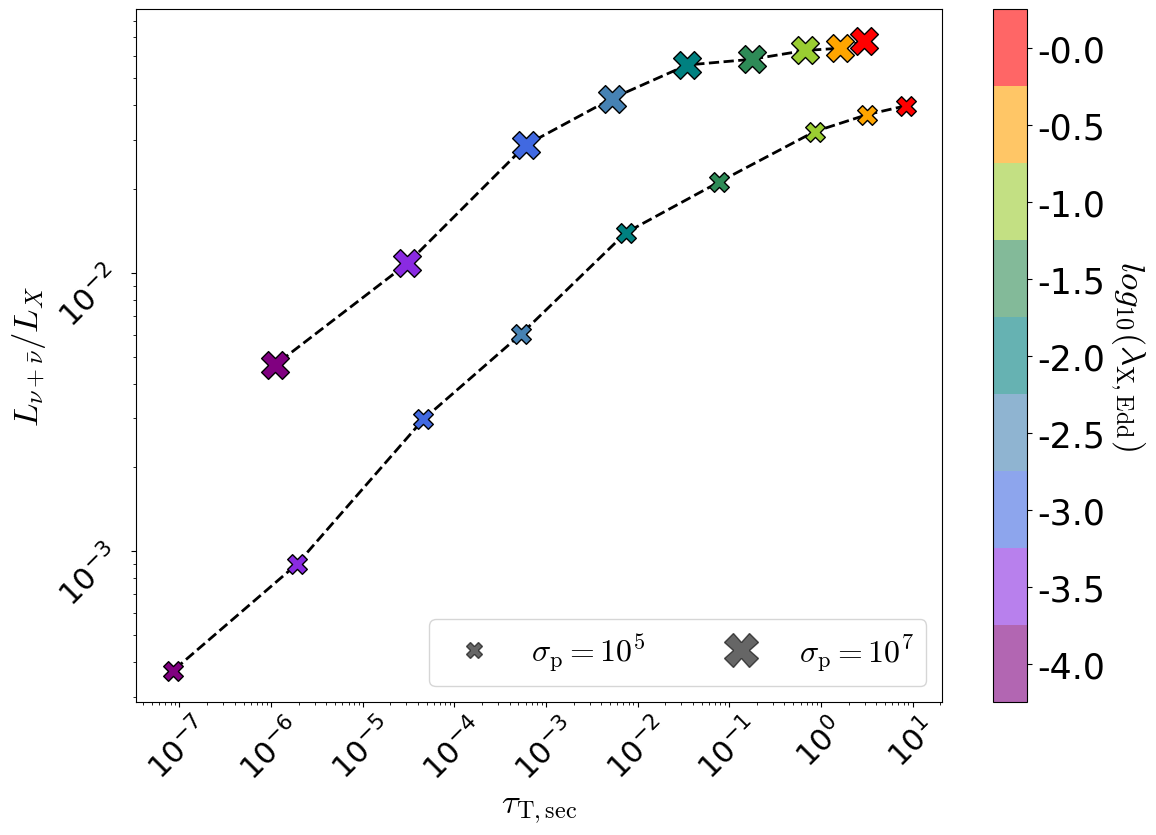}
    \caption{All-flavor (bolometric) neutrino luminosity over corona X-ray luminosity  as a function of Thomson optical depth due to secondary pairs for cases with $L_{\rm X}=10^{43}~erg~s^{-1}$ and varying black hole mass (see colorbar),  for two values of $\sigma_{\rm p}$ (see legend).}
    \label{fig:tau_T}
\end{figure}

Furthermore, the right-hand side panels of figure \ref{fig:ne_max} show that the fraction of the coronal luminosity transferred to neutrinos is proportional to $\lambda_{\rm X, Edd}$ for an optically thin source and becomes constant for an optically thick one, as described in detail in section \ref{sec:analytic-neutrino}. Moreover, the fraction of luminosity transferred to neutrinos is the same for sources with the same $\lambda_{\rm X, Edd}$ (and $\sigma_{\rm p}$, even though the black hole masses and broadband X-ray luminosities are different (see, for example, overlapping green squares with red crosses in right panels of figure \ref{fig:ne_max}). By comparing the upper and lower panels on the right-hand side of the figure we observe that the fraction of $L_{\rm X}$ transferred to neutrinos is higher for higher $\sigma_{\rm p}$ values (and fixed $\lambda_{\rm X, Edd}$). This trend can be understood as follows. 
For lower $\sigma_{\rm p}$ values, such as $10^5$, protons that carry most of the population energy, i.e. those with energies of $E_{\rm p, br}$, are close to threshold for Bethe-Heitler interactions with the coronal photons (see also proton timescale plot in figure~\ref{fig:timescales}). As a result, a non-negligible amount of the energy available in the proton population is channeled to secondary Bethe-Heitler pairs, leaving less energy available for neutrino production. The aforementioned condition is not taken into account in our analytical calculations (see instead Eq.~6 in Ref.~\cite{2021ApJ...906..131M} for a more accurate treatment), resulting in the over-prediction of $L_{{\nu}+\bar{\nu}}/L_{\rm X}$ for $\sigma_{\rm p}=10^5$ (see top right panel of figure \ref{fig:ne_max}). Nonetheless, the analytical predictions of the ratio (see Appendix~\ref{app:Lnu-bolo}) for the optically thin regime,  i.e. for $\lambda_{\rm X, Edd} \lesssim \lambda_{\rm Edd, crit}$, are in very good agreement with the numerical results, and with the analytical approximation as derived by \cite{2024ApJ...961L..14F} (solid black line).

We can combine the information shown in figure~\ref{fig:ne_max} into a diagram showing the bolometric neutrino luminosity over the broadband X-ray luminosity as a function of the Thomson optical depth attributed to secondary pairs (see figure \ref{fig:tau_T}). We observe that as $\lambda_{\rm X, Edd}$ increases (see colorbar) both the bolometric neutrino luminosity and the Thomson optical depth increase. Interestingly, for those cases with $0.1 \lesssim \tau_{\rm T, sec} \lesssim 10$, the bolometric neutrino luminosity is $\gtrsim 10^{-2} L_{\rm X}$, with the highest values expected when $\sigma_{\rm p}\sim 10^7$.

\section{Neutrino predictions for Seyfert galaxies} \label{sec:seyferts}
The coronal region of AGN has been proposed as a production site of high-energy neutrinos since the late seventies \citep[e.g.][]{1979ApJ...232..106E,1981MNRAS.194....3B}. The IceCube Collaboration has recently presented compelling evidence of high-energy neutrino emission (significance of 4.2$ \sigma$) from NGC~1068 \cite{IceCube-NGC1068}, a nearby Seyfert II galaxy located at a distance of $\sim$10.1 Mpc (for a review, see \cite{2024arXiv240520146P}). The best-fit neutrino spectrum, under the assumption of a power law distribution, is soft, $ d\Phi_\nu/dE_{\rm \nu}\propto E_{\rm \nu}^{-3.2}$, extending from 1.5 to 15 TeV. These experimental results can be reproduced by the model presented in section~\ref{sec:model}, as demonstrated in detail in Ref.~\cite{2024ApJ...961L..14F}. More recently, the IceCube Collaboration performed a stacking analysis of 27 additional Seyfert galaxies, which are part of the BAT AGN Spectroscopic Survey (BASS) \cite{2017ApJS..233...17R}, in search of a neutrino signal  \cite{2024arXiv240607601A}. They report excesses of neutrinos (at $ 2.7\sigma$ significance compared to the background expectation) from two Seyfert galaxies, NGC~4151 and CGCG~420-015. Furthermore, NGC~4945 is a close-by (located at 3.8 Mpc) obscured Seyfert II galaxy and a very bright emitter in the infrared energy regime, with that radiation assumed to be the reprocessed UV and optical emission originating from the center of the source, near the massive black hole. The aforementioned source, also, emits in the hard X-rays (14-195 keV), being comparably luminous to NGC~4151 \cite{aguilar-ruiz_cosmic_2021}. Moreover, NGC~4388 located at a distance of 21 Mpc is one of the closest and X-ray brightest Seyfert II galaxies. Also, NGC~6240 is an interesting source considering that it consists of two merging supermassive black holes that are at a distance less than 10 kpc from each other \cite{koss_population_2018}. This system, located approximately 100 Mpc from our Galaxy, is also observed by Chandra to emit 6-keV X-ray radiation \cite{wang_fast_2014}. Such characteristics make NGC~4945, NGC~4388 and NGC~6240 interesting targets for future observatories, such as IceCube-Gen2 \cite{Aartsen_2021} and KM3Net \cite{Adrián-Martínez_2016}.

\subsection{Application on individual sources}

We apply the model as described in section \ref{sec:model} with the difference that the maximum proton energy is not fixed at $10^8~\rm GeV$. Instead, the latter is set by the minimum value between the Hillas criterion \cite{Hillas_1984}, and the equilibrium between the characteristic acceleration time and the characteristic cooling time (see, for example, figure \ref{fig:timescales}). For NGC~1068, NGC~4151, CGCG~420-015, NGC~4954, NGC~4388 and NGC~6240 the maximum proton energies are: $1.8 \cdot 10^8$ GeV, $5.8 \cdot 10^{8}$ GeV, $5.8 \cdot 10^{8}$ GeV, $1.1 \cdot 10^{8}$ GeV, $2.9 \cdot 10^{8}$ GeV and $7.3 \cdot 10^{8}$ GeV, respectively.

\begin{table}[h]
\centering
\begin{threeparttable}
\centering
\caption{Properties of Seyfert galaxies used in our analysis.}
\begin{tabular}{l c c c c c c }
\hline
 Name    &  Distance & $L_{\rm 2-10 \, keV}$ & $L_{\rm X}^\dag$  & $M_{\rm bh}$ & $\lambda_{\rm X, Edd}$ & Refs.  \\   
 & (Mpc) & (erg s$^{-1}$) & (erg s$^{-1}$) & ($M_{\rm \odot}$)  \\  \hline
 NGC 1068 & 10.1 & $8.5 \cdot 10^{42}$ & $3.6\cdot 10^{43}$ & $6.7\cdot 10^6$ & $5 \cdot 10^{-2}$ &  [I] \\
 NGC 4151 &  15.8 & $2.0 \cdot 10^{42} $ & $8.8 \cdot 10^{42}$  &$3\cdot 10^7$~$^{\ddag}$ & $3 \cdot 10^{-3}$ & [II]  \\
 CGCG~420-015 & 128.8& $1.0 \cdot 10^{44}$  &  $4.3\cdot 10^{44}$ & $2\cdot 10^8$ & $2 \cdot 10^{-2}$ & [III] \\
 NGC 4945 & 3.8 &$1.2 \cdot 10^{42}$ &$5 \cdot 10^{42}$&$4 \cdot 10^6$& $3 \cdot 10^{-2}$ & [IV]\\
NGC 4388 & 21.0 &$1.1 \cdot 10^{43}$ &$4.8 \cdot 10^{43}$&$ 1.6 \cdot 10^7$& $3 \cdot 10^{-2}$ & [V]\\ 
NGC 6240$^{*}$& 106.9 &$ 5.6 \cdot 10^{44}$ &$ 2.4 \cdot 10^{45}$& $8 \cdot 10^8$& $3 \cdot 10^{-2}$ & [V]\\
\hline  
\end{tabular}
\begin{tablenotes}
\footnotesize{
\item $^\dag$ Based on the intrinsic 2-10 keV X-ray luminosity by \cite{ricci_bat_2017}
and recalculated to the 0.1-100 keV band assuming a photon index $s_X=2$.
\item $^\ddag$ Average value from different methods: $M_{\rm bh} = (3.76 \pm 1.15)\cdot 10^7 M_\odot$
\cite[from stellar dynamics,][]{2014ApJ...791...37O},  $M_{\rm bh} = 3.0^{+0.8}_{\rm -2.2}\cdot 10^7 M_\odot$
\cite[from gas dynamics,][]{2008ApJS..174...31H},  and $M_{\rm bh}=2.45^{+0.75}_{\rm -0.55}\cdot 10^7 M_{\rm \odot}$ \cite[from reverberation mapping,][]{2018ApJ...866..133D}.
\item $^{*}$ For the case of NGC~6240, since there are two supermassive black holes in the center of the galaxy, we use the values for the largest of the two and multiply the output of \code \ by a factor of 2.}\\
Parameter referencing: {\bf [I]:} distance and mass by \cite{2024arXiv240520146P} {\bf [II]:} mass by \cite{2018ApJ...866..133D} and distance by \cite{2020ApJ...902...26Y} {\bf [III]:} mass and distance by \cite{2017ApJ...850...74K} {\bf [IV]:} mass by \cite{greenhill_distribution_nodate} and distance by \cite{karachentsev_hubble_2007}
{\bf [V]}: mass by \cite{koss_bass_2022} and distance by \cite{2017ApJS..233...17R}.
\end{tablenotes}
\label{tab:seyf_gal_params}
\end{threeparttable}
\end{table}

\begin{figure}
    \centering
    \includegraphics[width=\textwidth]{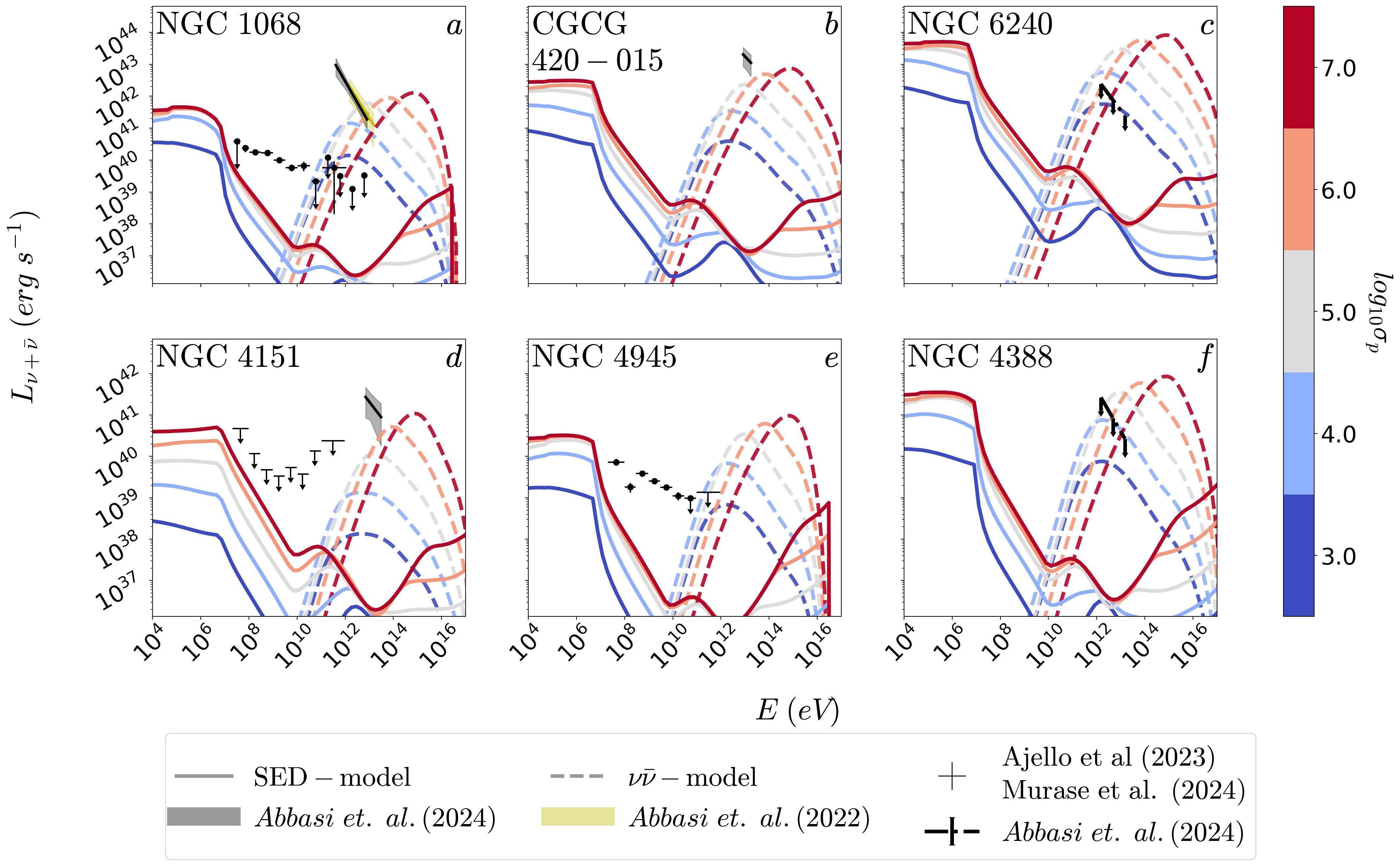}
    \caption{Numerical results of all-flavor neutrino spectra for NGC~1068 (panel a), CGCG~420-015 (panel b), NGC~6240 (panel c), NGC~4151 (panel d), NGC~4945 (panel e) and NGC~4388 (panel f), computed for a range of $ \sigma_{\rm p}$ values (see color bar). Solid lines represent the model cascade spectrum for each $\sigma_p$ value. Grey-shaded bands refer to the IceCube data by \cite{2024arXiv240607601A}, while the yellow area represent the best fit and 1$ \sigma$ deviation from Ref.~\cite{IceCube-NGC1068}. Dash-dotted black lines with arrows represent the upper limits as given by \cite{2024arXiv240607601A}. Black points represent Fermi-LAT and MAGIC observations and upper limits from Ref.~\cite{ajello_disentangling_2023} for NGC~1068 and Ref.~\cite{murase_sub-gev_2024} for NGC~4145 and NGC~4945.}
    \label{fig:seyf_gal_neutr}
\end{figure}

\begin{figure}
    \centering
    \includegraphics[width=0.8\linewidth]{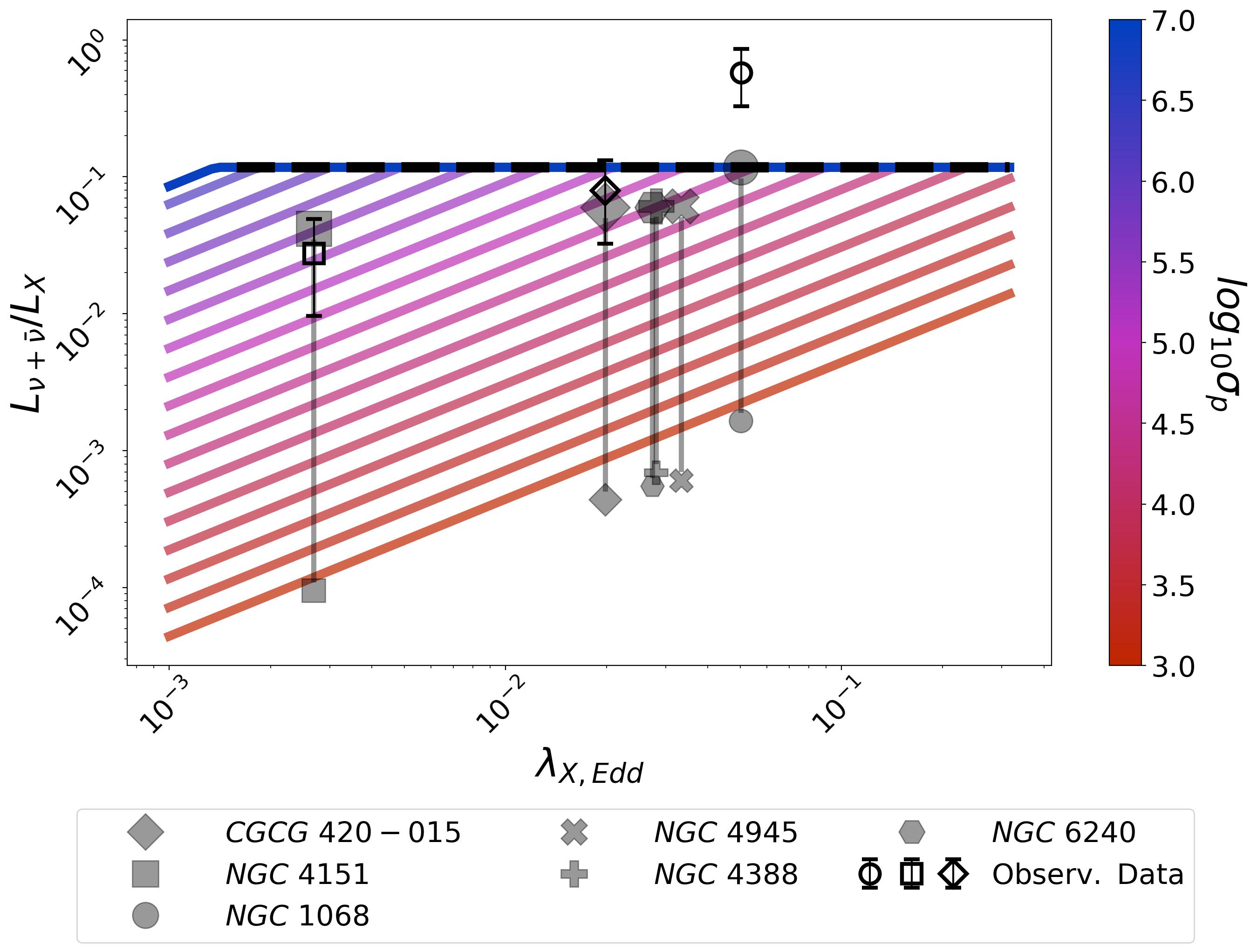}
    \caption{ Ratio of bolometric neutrino luminosity and broadband X-ray corona luminosity as a function of the Eddington ratio. Colored curves are computed using the analytical approximation of Eq. \ref{eq:nuLnu} estimated on the approximate peak neutrino energy, $ E_{\rm \nu, br}$, for various $ \sigma_{\rm p}$ values (see color bar). The horizontal dashed black line represents the calorimetric limit (Eq. \ref{eq:nuLnu}). The various filled markers represent the model predictions for the sources shown in figure \ref{fig:seyf_gal_neutr} for $\sigma_p = 10^3-10^7$, with the size of the marker increasing with higher $\sigma_p$ values. 
    The observational values (open markers) were acquired by \cite{2024arXiv240607601A} and Table \ref{tab:seyf_gal_params}.  The vertical bars on the open symbols are computed from the 1$ \sigma$ contours of the power-law fits presented in \cite{2024arXiv240607601A}, as explained in the text.}
    \label{fig:nuLnu_ratio_seyf}
\end{figure}

In figure \ref{fig:seyf_gal_neutr} we present the cascade photon spectra (above 10 keV) and the all-flavor neutrino spectra predicted by our model for the galaxies of Table \ref{tab:seyf_gal_params} and $ \sigma_{\rm p}= 10^3-10^7$. The latter is the main free parameter of the model for a given source. We overplot $\gamma$-ray flux measurements and upper limits, whenever available, and display the best-fit (all flavor) neutrino luminosity and the 68\% statistical uncertainty (grey-shaded bands) based on a power-law model as obtained by IceCube\footnote{We convert the reported energy fluxes to luminosities by using the luminosity distances listed in Table~\ref{tab:seyf_gal_params}.} {for three AGN}~\cite{2024arXiv240607601A}. We also overplot, whenever available, the IceCube upper flux limits (dash-dotted black lines with arrows) \citep{2024arXiv240607601A}. The shape of the unfolded neutrino data critically depends on the model used in the IceCube likelihood analysis, as can be seen in figure 3 in Ref.~\cite{2024arXiv240607601A}. As a result, a direct comparison of our model spectra to the unfolded power-law IceCube spectra would be ill-considered. For this reason, instead of fitting our model to the unfolded IceCube spectra, we focus on qualitatively assessing whether the neutrino signal predicted by the model can approximately account for the observed data.

Because the model neutrino spectra peak at higher energies as the proton magnetization increases, $\sigma_p$ values much larger than $10^5$ are disfavored for the three AGN with available neutrino data. Furthermore, our model with $\sigma_p=10^5$ is the closest one to the neutrino observations\footnote{An accurate determination of $ \sigma_{\rm p}$ would require fitting the IceCube neutrino events under the assumption that the signal is described by our model, a procedure that lies beyond the scope of this paper.} of NGC~1068. This result is in agreement with the findings of Ref.~\cite{2024ApJ...961L..14F}. Minor differences between the two model spectra arise due to slightly different values used for $\sigma_{\rm p}$ and $\eta_{\rm p}$ (in Ref.~\cite{2024ApJ...961L..14F} these were fine-tuned to better match the minimum energy and flux of the IceCube power-law neutrino spectrum). However, for NGC~4151 and CGCG~420-015, the model prediction for $\sigma_p=10^5-10^6$ underpredicts the observed luminosity. Although a higher value of $\eta_{\rm p}$ would reduce this discrepancy, it cannot be arbitrarily high because of the physical requirement that $\eta_{\rm X} + \eta_{\rm p} \le 1$. Finally, the cascade photon spectra for all $\sigma_p$ values we considered are well below the available $\gamma$-ray measurements (or upper limits). Therefore, in a reconnection-powered corona, the GeV $\gamma$-ray emission of NGC~1068 and NGC~4945 cannot have a coronal origin, in agreement with other studies~\citep[e.g.][]{murase_hidden_2020, inoue_origin_2020}. For NGC~4388 and NGC~6240, the model yields neutrino spectra less luminous than the upper limits calculated by the power-law model in \cite{2024arXiv240607601A} for $\sigma_p\lesssim 10^5$ and $\sigma_p \lesssim 10^4$ respectively.

In figure \ref{fig:nuLnu_ratio_seyf} we show the ratio of the bolometricneutrino luminosity to the broadband X-ray luminosity, as produced by our model filled markers), and in comparison to observational data for NGC~1068, NGC~4151 and CGCG~420-015  (open markers). The observational values were computed by integrating over energies the best-fit power-law model for the neutrino flux, $ \Phi(E_{\rm \nu}) \propto E^{-\gamma}$, acquired by \cite{2024arXiv240607601A}, and using the luminosity distances and black hole masses listed in Table \ref{tab:seyf_gal_params}. The vertical bars on the open symbols indicate the range of minimum and maximum fluxes obtained when using the 1$ \sigma$ contours of the power-law fits presented in figure 8 of Ref.~\cite{2024arXiv240607601A}. The model predictions were calculated for $\sigma_p$ starting from $10^3$ (lowest ratio) and reaching to $10^7$ (highest ratio).
Moreover, the colored curves in figure \ref{fig:nuLnu_ratio_seyf} represent the analytical approximation of Eq.~\ref{eq:nuLnu} for a coronal environment that is optically thin to $p\gamma$ interactions ($ f_{\rm p\gamma}<1$ -- see section \ref{sec:analytic} and Eq.~\ref{eq:ledd_cal_lim}), while the black dashed line describes the system when it is in the optically thick regime. We remind that all results were calculated for $\eta_{\rm p} = 1/3$  and $\eta_{\rm X}= 1/2$. If both were equal to 1/2, the ratio would increase to approximately 0.4 (see also Eq.~\ref{eq:Lnu_LX_thick}), moving closer to the NGC~1068 result (open circle). However, $\eta_{\rm p}$ and $\eta_{\rm X}$ are not expected to be vastly different, as this would mean large deviations from equipartition in the reconnection layer.

Summarizing, if one can observationally determine $ L_{\rm X}$, $M_{\rm bh}$, and $L_{\nu + \bar{\nu}}$, one can use figure~\ref{fig:nuLnu_ratio_seyf} and Eq.~\ref{eq:nuLnu} as guidelines to infer the proton magnetization in the coronal environment. Because $ \sigma_{\rm p}$ also relates to the peak energy of the neutrino spectrum, as shown in figure~\ref{fig:seyf_gal_neutr}, the inferred $ \sigma_{\rm p}$ value can be then cross-checked against neutrino spectra, whenever available.

\subsection{Stacked neutrino flux} \label{sec:mock_sample}

\begin{figure}
    \centering
    \includegraphics[width=0.8\linewidth]{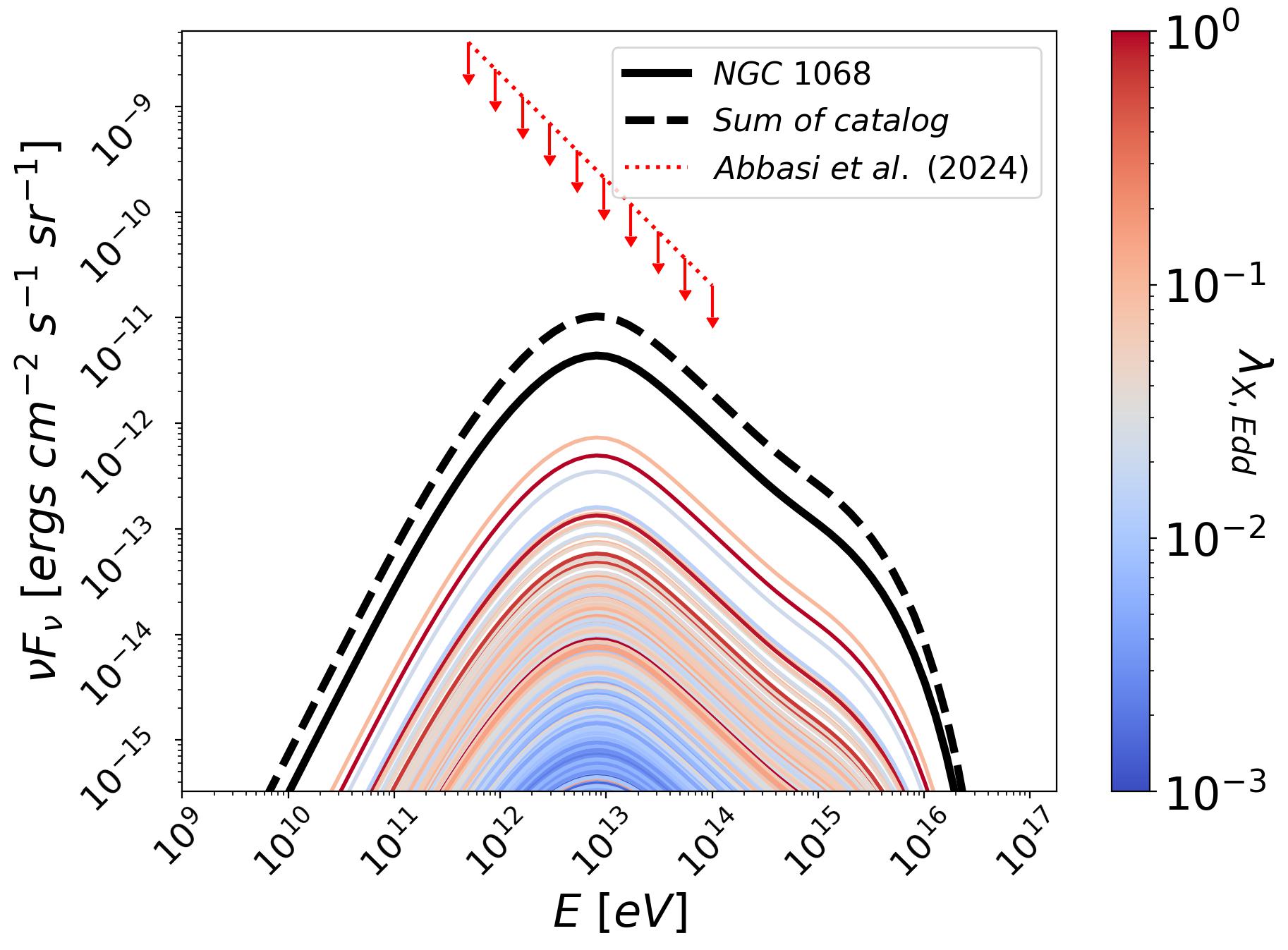}
    \caption{All-flavor neutrino energy flux of non-blazar AGN in the catalog of \cite{ricci_bat_2017} assuming similar spectral shape as NGC~1068 (black solid line) but different normalization that scales with their intrinsic X-ray luminosity. Colors represent the ratio $\lambda_{\rm X, Edd}$. The dashed black line shows the sum of all sources while the red dotted line with downward pointing arrows represents the neutrino flux upper limits for non-blazar sources from Table 1 of \cite{abbasi_search_2024}.}
    \label{fig:mock_sample}
\end{figure}

In this section we estimate the total neutrino flux expected from the 677 non-blazar AGN (including NGC~1068) of the Swift-BAT 70-month catalog \cite{ricci_bat_2017} based on our corona model. We assume that the neutrino spectrum for each source has the same shape as the one of NGC~1068, but different normalization. This is scaled with the X-ray intrinsic luminosity of the AGN, as given in the aforementioned catalog, and the ratio of the produced neutrino luminosity to the intrinsic X-ray luminosity using Eq.~\ref{eq:nuLnu}:

\begin{gather}
    \nu F_{\nu, i} = \nu L_{\nu, 1068}\frac{L_{X, i}}{L_{X, 1068}} \frac{\nu L_{\nu, i}}{L_{X, i}} \left( 4 \pi D_i \right)^{-2}
    % \Rightarrow \nu F_{\nu, i}= \nu L_{\nu, i} \frac{\nu L_{\nu, 1068}}{L_{X, 1068}} \left( 4 \pi D \right)^{-2} 
    \label{eq:mock_sample}
\end{gather}
where $D$ is the source distance obtained from the catalog, and the subscripts $i$ and $1068$ are used to indicate the $i-$th galaxy from the sample and NGC~1068, respectively. The results of the analysis described above are shown in figure \ref{fig:mock_sample}, where the black solid line represents the spectrum of NGC~1068, as predicted by our model, and the colored solid lines show the fluxes of each of the other 676 sources. Moreover, the black dashed line is the sum of all 677 sources, including NGC~1068, while the red dotted line represents the neutrino flux upper limits for non-blazar sources given by Table 1 in \cite{abbasi_search_2024}.

In figure \ref{fig:mock_sample}, we see that NGC~1068 is among the brightest contributors to the stacked neutrino flux. The total flux predicted by our model, assuming that all the non-blazar sources produce an NGC~1068 like spectrum, is consistent with the upper limits calculated in \cite{abbasi_search_2024}. However, in future work, we plan on making a more detailed calculation of the diffuse neutrino flux by combining our model with a luminosity function for non-blazar sources. We will also consider an intrinsic distribution of corona magnetizations, resulting in the creation of morphologically different individual spectra.

\section{Summary and Discussion} \label{sec:discuss}
We introduced a model for reconnection-powered AGN coronae. We assumed that a fraction of the dissipated magnetic energy via reconnection is used to power the X-ray coronal emission, whose spectrum was described by a power law between 0.1 keV and 100 keV with photon index 2. We also considered that protons existing in the coronal environment are accelerated in magnetic reconnection regions near the central black hole. The number density distribution of the injected hadrons into the corona is described by a broken power law, with slopes of $-1$ and $-3$ respectively before and after the break energy $ \sigma_{\rm p} m_{\rm p} c^2$, where $ \sigma_{\rm p}$ is the proton plasma magnetization in the corona. These relativistic protons interacting with coronal photons produce secondary pairs and high-energy neutrinos. The pairs can contribute to enrich the lepton population in the corona, while the neutrinos can provide a novel signal visible at Earth. 
Based on the aforementioned model, we performed a parameter scan of the coronal X-ray luminosity, black hole mass, and proton plasma magnetization. For each set of parameters, we computed the high-energy neutrino spectrum and estimated the number density of secondary pairs produced in the corona using the leptohadronic code \code. We complemented our parametric study with analytical calculations that showcase the impact of the three main physical parameters.

Describing the differential photon density spectrum of the corona as a power law with slope $-2$ can be considered a simplification which, however, does not significantly alter our results. In Ref.~\cite{trakhtenbrot_bat_2017}, the authors found, by fitting observational data of 228 AGN, that the coronal photon index, $s_{\rm X}$ is proportional to $\log_{10}(L_{\rm bol}/ L_{\rm Edd})$, with $L_{\rm bol}$ being the bolometric AGN luminosity. The aforementioned dependence can be considered a weak one and, thus, the parameter space that creates for the $s_{\rm X}$ is very narrow, with $s_{\rm X} \in [1, 3]$. For $s_{\rm X}>1$, most of the coronal photons, in number, are found at the lower limit of the distribution (e.g. 0.1 keV) and are the dominant targets for photo-hadronic interactions. As a result, changes in $s_{\rm X}$ would not affect the overall picture of the corona apart from its luminosity; this can be altered, at most, by a factor of 9 compared to the one obtained for the nominal value $s_{\rm X}=2$.

Our results, both numerical and analytical, highlight the importance of the Eddington ratio parameter, $\lambda_{\rm X, Edd}$. Such a dependence results in the initially 3D parameter space of our model ($ L_{\rm X}$, $ M_{\rm bh}$, $ \sigma_{\rm p}$) being reduced to a 2D parameter space ($\lambda_{\rm X, Edd}$, $ \sigma_{\rm p}$), with the X-ray luminosity $L_{\rm X}$ only fixing the overall energetics of the source. More specifically, we find that for $\lambda_{\rm X, Edd} \lesssim \lambda_{\rm Edd, crit}$ (see Eq.~\ref{eq:ledd_cal_lim}) the system is optically thin to $p\gamma$ interactions. As a result, the latter are not capable of sustaining a high-enough pair density in the corona to account for the inferred Thomson scattering depths of AGN coronae. Therefore, for $\lambda_{\rm X, Edd} \lesssim \lambda_{\rm Edd, crit}$, the percentage of the coronal luminosity, $L_{\rm X}$, transferred to neutrinos is proportional to the Eddington ratio, $\lambda_{\rm X, Edd}$, while that aforementioned percentage becomes constant otherwise (see e.g. Eq.~\ref{eq:Lnu_LX_thick} and figure~\ref{fig:ne_max}), corresponding to a calorimetric regime where all of the protons dump their energy in secondary particles.

In this work we have not specified the origin of the pairs responsible for the Comptonization of soft photons to the X-ray regime. Nonetheless, the pair number density in the corona is expected to be the sum of hadronic-generated (secondary) pairs and of primary pairs (i.e., those generated by other channels). For the parametric search we performed in section~\ref{sec:res} we found  $ 10^{-7} \lesssim \tau_{\rm T, sec} \lesssim 10$, with the highest values obtained in systems with the lightest black holes and highest proton plasma magnetizations (see figure~\ref{fig:tau_T}). Assuming that the total Thomson optical depth (with both primary and secondary contributions) is $ \tau_T \sim 1$, we estimated the pair plasma magnetization for the cases we explored (see e.g. Eq.~\ref{eq:sigma_e_Edd}) and found that $ 10^{-5} \lesssim  \sigma_{\rm e}/\sigma_{\rm p} \lesssim 1$.

Moreover, we comment on the dependence of our results on the post-break slope $s_{\rm p}$ of the accelerated proton population, which was taken to be 3 so far. 3D PIC simulations of reconnection in the relativistic regime $ 1 \ll \sigma_e < \sigma_{\rm p}$, which is of relevance to our work, have not yet been performed. There is evidence, however, that the post-break slope is related to the strength of the non-reconnecting magnetic field, known as the guide field. 3D PIC simulations in electron-positron plasmas indicate steeper power laws for stronger guide fields (see e.g. Figs.~3 and 4 in \cite{2017ApJ...843L..27W}). For $ s_{\rm p}=2$ the energy of the hadronic population is equally distributed in logarithmic energy to protons with $ E_{\rm p} \geq E_{\rm p, br}= \sigma_{\rm p} m_{\rm p} c^2$. As a result, all protons above the break play an important part in pair creation and neutrino production. We performed numerical runs for the same parameters as those used in figure \ref{fig:spectr_neutr_prot} except for $ s_{\rm p}=2$ (Appendix~\ref{App:slope_2}). We find that, in this case, the change of $ \sigma_{\rm p}$ does not significantly affect the shape and luminosity of the electromagnetic spectrum or the luminosity and peak position of the neutrino energy distribution (see figure \ref{fig:sp_2}).

In this study we assumed that the X-ray non-thermal emission of the corona is fixed (i.e., it does not depend on the hadronic interactions and does not evolve with time). Under this hypothesis, we computed the expected density of secondary pairs created by proton-photon  and photon-photon interactions. In cases where hadronic interactions are responsible for producing a high density of secondary pairs, the assumption of a fixed coronal field has to be modified, and a time-dependent framework should be adopted. We envision a situation where accelerated protons start interacting with an ambient soft thermal radiation field, e.g. the UV bump of the accretion disk, and produce secondary pairs which will contribute to the pool of non-relativistic pairs in the corona. Such pairs can Comptonize soft photons to higher energies. As a result, a fraction of the injected energy into secondary pairs can be transferred to X-ray photons, that will obtain a power-law spectrum dictated by the pair density in the corona \cite{Lightman}. Therefore, protons will be interacting with an evolving radiation field. In this new description of the coronal environment, it is not clear whether a steady state can be reached, and if its emission will be characterized by a hard (photon index $ \lesssim 2$ ) non-thermal X-ray photon field. We intend to improve our model by accounting for an evolving coronal radiation field that will be dynamically coupled to the relativistic proton population of the corona in a separate work.

We have also performed some numerical runs for XRB-like environments, with the mass of the compact object being $ M_{\rm bh}=1-10 ~M_{\rm \odot}$ and a broadband X-ray luminosity of $ L_{\rm X}=10^{37}-10^{38} \rm ~erg ~s^{-1}$. The aforementioned parameters yield $\lambda_{\rm X, Edd}$ values equivalent to the AGN case of $ L_{\rm X}=10^{43}\rm ~erg~ s^{-1}$ and $ M_{\rm bh}=10^6~M_{\rm \odot}$ (discussed in Section \ref{sec:res}). We find that such XRB-like systems behave in a similar way as the AGN studied in this work. More specifically, electromagnetic and neutrino spectra are shaped like the ones presented in sections \ref{sec:res} and \ref{sec:seyferts}, but they are less luminous than those of AGN coronae; their luminosities scale with the X-ray luminosity of the system. We attribute the similarities to the fact that both XRB and AGN corona we studied have similar compactnesses (in terms of photons and protons).

In this work, as mentioned above, we consider particle energization in reconnection environments. Other studies, on the other hand, such as \cite{murase_hidden_2020, fiorillo_magnetized_2024} consider that particles are accelerated in turbulent environments. Although both reconnection and turbulence models describe the corona as a magnetized region, there are some fundamental differences between the two scenarios. For instance, magnetic reconnection is considered to take place in environments with $n_e \gg n_p$ resulting in $\sigma_e \ll \sigma_p$, with $\sigma_p \gg 1$. Additionally, in reconnecting plasma environments, the acceleration timescale of particles is taken to be proportional to their Lorentz factor, in reconnecting plasma environments. PIC simulations have shown that ions can reach Lorentz factor values of the order of proton magnetization $\rm \sigma_p$ \cite{com2024}. In contrast, in the turbulence models, the corona is associated with the accretion flow, resulting in the assumption that $n_{\rm p} \simeq n_{\rm e}$ and so the magnetization of the system is of the order of unity or less \cite{fiorillo_magnetized_2024}. In such models the acceleration is diffusive and its timescale may \cite{murase_hidden_2020} or may not \cite{fiorillo_magnetized_2024} depend on the proton Lorentz factor. For example, in Ref.~\cite{murase_hidden_2020} a power-law in energy, with slope linked to the turbulent power spectrum, is used to describe the acceleration timescale, whilst in reconnection (see also \cite{2024ApJ...961L..14F}) it depends linearly on the proton energy (see figure \ref{fig:timescales}). Moreover, reconnection models assume coronal environments of a few $r_{\rm  g}$ (see also \cite{2024ApJ...961L..14F}) while in turbulence models the size of the corona can reach up to 20 or even 100 $r_{\rm g}$ \cite{murase_hidden_2020, eichmann_solving_2022}. The higher values of the magnetic field in reconnection models, also, make the contribution of meson synchrotron losses significant, as opposed to the turbulent scenario. In the latter, the Bethe-Heitler losses on the OUV photons can also be more important and p-p losses become relevant, as the relativistic protons are in a region with dense plasma, as opposed to the reconnection environments. The differences outlined above lead to different shapes of the accelerated proton distribution, which, in turn, are imprinted on the neutrino spectra.

Furthermore, in such highly magnetized environments such those of reconnection-powered coronae, synchrotron radiation from heavier particles, such as pions or mesons might become important-- such effects have been studied in GRB environments (see \cite{waxman_high_1997, petropoulou_implications_2014, florou_marginally_2021}). Even though pions and muons have short decay timescales, if the magnetic field is significantly strong they might be able to radiate a fraction of their energy before they decay (see Appendix \ref{App:cooling}), affecting the shape of the produced neutrino spectrum. In particular for low $\sigma_p$ values, while the peak position and luminosity of the neutrino energy spectrum remain unchanged, the high-energy spectrum will appear steeper compared to spectra where no meson cooling was taken into account (right panel of figure \ref{fig:seyf_gal_neutr_heavy}). For even higher $\sigma_p$ values (e.g. $10^6-10^7$), meson cooling due to synchrotron can also affect the luminosity and position of the neutrino maximum (left panel of figure \ref{fig:seyf_gal_neutr_heavy}). The aforementioned steepening of the neutrino spectrum is also visible in neutrino spectra produced by protons with post-break slope of $s_p=2$ as described in Appendix \ref{App:slope_2}. As a result the neutrino spectra shown in the right panel of figure~\ref{fig:sp_2} will become softer, better matching the observed IceCube spectra of NGC~1068 or CGCG~420-015.

An interesting aspect of our model is that the neutrino spectrum from an AGN corona can be determined if basically two parameters are known, the ratio $\lambda_{\rm X, Edd}$ and $ \sigma_{\rm p}$ (if $ s_{\rm p}$ is specified). The constraining nature of the model and its simplicity make it appealing for the calculation of the diffuse neutrino flux from Seyfert galaxies. Contrary to other works where NGC~1068 is used as a prototype, our model suggests that NGC~1068 may not be representative of the population in terms of neutrino production (see e.g. figure~\ref{fig:nuLnu_ratio_seyf}). In fact, we find that our model can satisfactorily represent the neutrino signal inferred from NGC~1068, CGCG~420-015 and NGC 4151 by IceCube requiring the latter two sources having larger $ \sigma_{\rm p}$ values than NGC~1068.

Our two-parameter model can be extended to predict the neutrino flux from the AGN corona population. Being a superposition of sources with different properties (i.e., $\lambda_{\rm X, Edd}$ and $ \sigma_{\rm p}$), the diffuse neutrino spectrum of the population is expected to be broader than the one computed using a single source as a prototype (see Sec. \ref{sec:mock_sample} and \cite{2024A&A...684L..21P} and \cite{2017MNRAS.470.1881P} in a different context). A comparison of our model prediction against the most updated IceCube measurement of the diffuse flux will be studied in a separate work.

\acknowledgments
We thank the anonymous referee for their insightful comments that helped us clarify several points in the manuscript. We also thank Chengchao Yuan for his useful comments on this manuscript. M.P. acknowledges support from the Hellenic Foundation for Research and Innovation (H.F.R.I.) under the "2nd call for H.F.R.I. Research Projects to support Faculty members and Researchers" through the project UNTRAPHOB (Project ID 3013). DFGF is supported by the Alexander von Humboldt Foundation (Germany) and, when this work was started, was supported by the Villum Fonden (Denmark) under Project No. 29388 and the European Union’s Horizon 2020 Research and Innovation Program under the Marie Sklodowska-Curie Grant Agreement No. 847523 'INTERACTIONS'. L.C. acknowledges support from NSF grant PHY-2308944 and NASA ATP grant 80NSSC22K0667. L.S. acknowledges support from DoE Early Career Award DE-SC0023015 and NASA ATP 80NSSC24K1238. This work was also supported by a grant from the Simons Foundation (MP-SCMPS-00001470) to L.S., and facilitated by Multimessenger Plasma Physics Center (MPPC, NSF PHY-2206609 to L.S.).

\bibliographystyle{JHEP}
\bibliography{bibliography}

\providecommand{\href}[2]{#2}\begingroup\raggedright\begin{thebibliography}{10}

\bibitem{2017MNRAS.465..358C}
J.S.~{Collinson}, M.J.~{Ward}, H.~{Landt}, C.~{Done}, M.~{Elvis} and J.C.~{McDowell}, \emph{{Reaching the peak of the quasar spectral energy distribution - II. Exploring the accretion disc, dusty torus and host galaxy}}, \href{https://doi.org/10.1093/mnras/stw2666}{\emph{\mnras} {\bfseries 465} (2017) 358} [\href{https://arxiv.org/abs/1610.04221}{{\ttfamily 1610.04221}}].

\bibitem{1980A&A....86..121S}
R.A.~{Sunyaev} and L.G.~{Titarchuk}, \emph{{Comptonization of X-Rays in Plasma Clouds - Typical Radiation Spectra}}, {\emph{\aap} {\bfseries 86} (1980) 121}.

\bibitem{1991ApJ...380L..51H}
F.~{Haardt} and L.~{Maraschi}, \emph{{A Two-Phase Model for the X-Ray Emission from Seyfert Galaxies}}, \href{https://doi.org/10.1086/186171}{\emph{\apjl} {\bfseries 380} (1991) L51}.

\bibitem{1999ApJ...510L.123B}
A.M.~{Beloborodov}, \emph{{Plasma Ejection from Magnetic Flares and the X-Ray Spectrum of Cygnus X-1}}, \href{https://doi.org/10.1086/311810}{\emph{\apjl} {\bfseries 510} (1999) L123} [\href{https://arxiv.org/abs/astro-ph/9809383}{{\ttfamily astro-ph/9809383}}].

\bibitem{2002ApJ...572L.173L}
B.F.~{Liu}, S.~{Mineshige} and K.~{Shibata}, \emph{{A Simple Model for a Magnetic Reconnection-heated Corona}}, \href{https://doi.org/10.1086/341877}{\emph{\apjl} {\bfseries 572} (2002) L173} [\href{https://arxiv.org/abs/astro-ph/0205257}{{\ttfamily astro-ph/0205257}}].

\bibitem{2017ApJ...850..141B}
A.M.~{Beloborodov}, \emph{{Radiative Magnetic Reconnection Near Accreting Black Holes}}, \href{https://doi.org/10.3847/1538-4357/aa8f4f}{\emph{\apj} {\bfseries 850} (2017) 141} [\href{https://arxiv.org/abs/1701.02847}{{\ttfamily 1701.02847}}].

\bibitem{2020ApJ...899...52S}
L.~{Sironi} and A.M.~{Beloborodov}, \emph{{Kinetic Simulations of Radiative Magnetic Reconnection in the Coronae of Accreting Black Holes}}, \href{https://doi.org/10.3847/1538-4357/aba622}{\emph{\apj} {\bfseries 899} (2020) 52} [\href{https://arxiv.org/abs/1908.08138}{{\ttfamily 1908.08138}}].

\bibitem{2021MNRAS.507.5625S}
N.~{Sridhar}, L.~{Sironi} and A.M.~{Beloborodov}, \emph{{Comptonization by reconnection plasmoids in black hole coronae I: Magnetically dominated pair plasma}}, \href{https://doi.org/10.1093/mnras/stab2534}{\emph{\mnras} {\bfseries 507} (2021) 5625} [\href{https://arxiv.org/abs/2107.00263}{{\ttfamily 2107.00263}}].

\bibitem{sridhar_comptonization_2022}
N.~Sridhar, L.~Sironi and A.M.~Beloborodov, \emph{Comptonization by reconnection plasmoids in black hole coronae {II}: {Electron}-ion plasma}, \href{https://doi.org/10.1093/mnras/stac2730}{\emph{Monthly Notices of the Royal Astronomical Society} {\bfseries 518} (2022) 1301}.

\bibitem{murase_hidden_2020}
K.~Murase, S.S.~Kimura and P.~Mészáros, \emph{Hidden {Cores} of {Active} {Galactic} {Nuclei} as the {Origin} of {Medium}-{Energy} {Neutrinos}: {Critical} {Tests} with the {MeV} {Gamma}-{Ray} {Connection}}, \href{https://doi.org/10.1103/PhysRevLett.125.011101}{\emph{Physical Review Letters} {\bfseries 125} (2020) 011101}.

\bibitem{kheirandish_high-energy_2021}
A.~Kheirandish, K.~Murase and S.S.~Kimura, \emph{High-energy {Neutrinos} from {Magnetized} {Coronae} of {Active} {Galactic} {Nuclei} and {Prospects} for {Identification} of {Seyfert} {Galaxies} and {Quasars} in {Neutrino} {Telescopes}}, \href{https://doi.org/10.3847/1538-4357/ac1c77}{\emph{The Astrophysical Journal} {\bfseries 922} (2021) 45}.

\bibitem{murase_high-energy_2023}
K.~Murase and F.W.~Stecker, \emph{High-{Energy} {Neutrinos} from {Active} {Galactic} {Nuclei}},  pp.~483--540 (2023), \href{https://doi.org/10.1142/9789811282645_0010}{DOI}.

\bibitem{2022Sci...378..538I}
{IceCube Collaboration}, R.~{Abbasi}, M.~{Ackermann}, J.~{Adams}, J.A.~{Aguilar}, M.~{Ahlers} et~al., \emph{{Evidence for neutrino emission from the nearby active galaxy NGC 1068}}, \href{https://doi.org/10.1126/science.abg3395}{\emph{Science} {\bfseries 378} (2022) 538} [\href{https://arxiv.org/abs/2211.09972}{{\ttfamily 2211.09972}}].

\bibitem{inoue_origin_2020}
Y.~Inoue, D.~Khangulyan and A.~Doi, \emph{On the {Origin} of {High}-energy {Neutrinos} from {NGC} 1068: {The} {Role} of {Nonthermal} {Coronal} {Activity}}, \href{https://doi.org/10.3847/2041-8213/ab7661}{\emph{The Astrophysical Journal Letters} {\bfseries 891} (2020) L33}.

\bibitem{eichmann_solving_2022}
B.~Eichmann, F.~Oikonomou, S.~Salvatore, R.-J.~Dettmar and J.B.~Tjus, \emph{Solving the {Multimessenger} {Puzzle} of the {AGN}-starburst {Composite} {Galaxy} {NGC} 1068}, \href{https://doi.org/10.3847/1538-4357/ac9588}{\emph{The Astrophysical Journal} {\bfseries 939} (2022) 43}.

\bibitem{MAGIC-UL-NGC1068}
V.A.~{Acciari} et~al., \emph{{Constraints on Gamma-Ray and Neutrino Emission from NGC 1068 with the MAGIC Telescopes}}, \href{https://doi.org/10.3847/1538-4357/ab3a51}{\emph{\apj} {\bfseries 883} (2019) 135} [\href{https://arxiv.org/abs/1906.10954}{{\ttfamily 1906.10954}}].

\bibitem{2022ApJ...941L..17M}
K.~{Murase}, \emph{{Hidden Hearts of Neutrino Active Galaxies}}, \href{https://doi.org/10.3847/2041-8213/aca53c}{\emph{\apjl} {\bfseries 941} (2022) L17} [\href{https://arxiv.org/abs/2211.04460}{{\ttfamily 2211.04460}}].

\bibitem{2024ApJ...961L..14F}
D.F.G.~{Fiorillo}, M.~{Petropoulou}, L.~{Comisso}, E.~{Peretti} and L.~{Sironi}, \emph{{TeV Neutrinos and Hard X-Rays from Relativistic Reconnection in the Corona of NGC 1068}}, \href{https://doi.org/10.3847/2041-8213/ad192b}{\emph{\apjl} {\bfseries 961} (2024) L14} [\href{https://arxiv.org/abs/2310.18254}{{\ttfamily 2310.18254}}].

\bibitem{Ripperda_2020}
B.~Ripperda, F.~Bacchini and A.A.~Philippov, \emph{Magnetic reconnection and hot spot formation in black hole accretion disks}, \href{https://doi.org/10.3847/1538-4357/ababab}{\emph{The Astrophysical Journal} {\bfseries 900} (2020) 100}.

\bibitem{el_mellah_reconnection-driven_2023}
I.~El~Mellah, B.~Cerutti and B.~Crinquand, \emph{Reconnection-driven flares in {3D} black hole magnetospheres: {A} scenario for hot spots around {Sagittarius} {A}*}, \href{https://doi.org/10.1051/0004-6361/202346781}{\emph{Astronomy \& Astrophysics} {\bfseries 677} (2023) A67}.

\bibitem{nathanail_magnetic_2022}
A.~Nathanail, V.~Mpisketzis, O.~Porth, C.M.~Fromm and L.~Rezzolla, \emph{Magnetic reconnection and plasmoid formation in three-dimensional accretion flows around black holes}, \href{https://doi.org/10.1093/mnras/stac1118}{\emph{Monthly Notices of the Royal Astronomical Society} {\bfseries 513} (2022) 4267}.

\bibitem{sironi_relativistic_2015}
L.~Sironi, M.~Petropoulou and D.~Giannios, \emph{Relativistic jets shine through shocks or magnetic reconnection?}, \href{https://doi.org/10.1093/mnras/stv641}{\emph{Monthly Notices of the Royal Astronomical Society} {\bfseries 450} (2015) 183}.

\bibitem{2019ApJ...880...37P}
M.~{Petropoulou}, L.~{Sironi}, A.~{Spitkovsky} and D.~{Giannios}, \emph{{Relativistic Magnetic Reconnection in Electron-Positron-Proton Plasmas: Implications for Jets of Active Galactic Nuclei}}, \href{https://doi.org/10.3847/1538-4357/ab287a}{\emph{\apj} {\bfseries 880} (2019) 37} [\href{https://arxiv.org/abs/1906.03297}{{\ttfamily 1906.03297}}].

\bibitem{chernoglazov_high-energy_2023}
A.~Chernoglazov, H.~Hakobyan and A.A.~Philippov, \emph{High-{Energy} {Radiation} and {Ion} {Acceleration} in {Three}-dimensional {Relativistic} {Magnetic} {Reconnection} with {Strong} {Synchrotron} {Cooling}},  Oct., 2023.

\bibitem{zhang_origin_2023}
H.~Zhang, L.~Sironi, D.~Giannios and M.~Petropoulou, \emph{The origin of power-law spectra in relativistic magnetic reconnection},  Feb., 2023.

\bibitem{zhang_fast_2021}
H.~Zhang, L.~Sironi and D.~Giannios, \emph{Fast particle acceleration in three-dimensional relativistic reconnection}, \href{https://doi.org/10.3847/1538-4357/ac2e08}{\emph{The Astrophysical Journal} {\bfseries 922} (2021) 261}.

\bibitem{2018MNRAS.480.1819R}
C.~{Ricci}, L.C.~{Ho}, A.C.~{Fabian}, B.~{Trakhtenbrot}, M.J.~{Koss}, Y.~{Ueda} et~al., \emph{{BAT AGN Spectroscopic Survey - XII. The relation between coronal properties of active galactic nuclei and the Eddington ratio}}, \href{https://doi.org/10.1093/mnras/sty1879}{\emph{\mnras} {\bfseries 480} (2018) 1819} [\href{https://arxiv.org/abs/1809.04076}{{\ttfamily 1809.04076}}].

\bibitem{2020A&A...634A..85P}
P.O.~{Petrucci}, D.~{Gronkiewicz}, A.~{Rozanska}, R.~{Belmont}, S.~{Bianchi}, B.~{Czerny} et~al., \emph{{Radiation spectra of warm and optically thick coronae in AGNs}}, \href{https://doi.org/10.1051/0004-6361/201937011}{\emph{\aap} {\bfseries 634} (2020) A85} [\href{https://arxiv.org/abs/2001.02026}{{\ttfamily 2001.02026}}].

\bibitem{2017ApJ...843L..27W}
G.R.~{Werner} and D.A.~{Uzdensky}, \emph{{Nonthermal Particle Acceleration in 3D Relativistic Magnetic Reconnection in Pair Plasma}}, \href{https://doi.org/10.3847/2041-8213/aa7892}{\emph{\apjl} {\bfseries 843} (2017) L27} [\href{https://arxiv.org/abs/1705.05507}{{\ttfamily 1705.05507}}].

\bibitem{com2024}
L.~{Comisso}, \emph{{Concurrent Particle Acceleration and Pitch-angle Anisotropy Driven by Magnetic Reconnection: Ion-electron Plasmas}}, \href{https://doi.org/10.3847/1538-4357/ad51fe}{\emph{\apj} {\bfseries 972} (2024) 9} [\href{https://arxiv.org/abs/2405.18227}{{\ttfamily 2405.18227}}].

\bibitem{Hillas_1984}
A.M.~{Hillas}, \emph{{The Origin of Ultra-High-Energy Cosmic Rays}}, \href{https://doi.org/10.1146/annurev.aa.22.090184.002233}{\emph{\araa} {\bfseries 22} (1984) 425}.

\bibitem{2003ApJ...586...79A}
A.M.~{Atoyan} and C.D.~{Dermer}, \emph{{Neutral Beams from Blazar Jets}}, \href{https://doi.org/10.1086/346261}{\emph{\apj} {\bfseries 586} (2003) 79} [\href{https://arxiv.org/abs/astro-ph/0209231}{{\ttfamily astro-ph/0209231}}].

\bibitem{karavola_BH_2024}
D.~Karavola and M.~Petropoulou, \emph{A closer look at the electromagnetic signatures of {Bethe}-{Heitler} pair production process in blazars},  May, 2024.

\bibitem{Dimitrakoudis}
S.~Dimitrakoudis, A.~Mastichiadis, R.J.~Protheroe and A.~Reimer, \emph{The time-dependent one-zone hadronic model - first principles}, \href{https://doi.org/10.1051/0004-6361/201219770}{\emph{A\&A} {\bfseries 546} (2012) A120}.

\bibitem{mastichiadis_spectral_2005}
A.~Mastichiadis, R.J.~Protheroe and J.G.~Kirk, \emph{Spectral and temporal signatures of ultrarelativistic protons in compact sources},  \href{https://arxiv.org/abs/astro-ph/0501156}{{\ttfamily astro-ph/0501156}}.

\bibitem{Petropoulou_2015}
M.~{Petropoulou}, S.~{Dimitrakoudis}, P.~{Padovani}, A.~{Mastichiadis} and E.~{Resconi}, \emph{{Photohadronic origin of {\ensuremath{\gamma}} -ray BL Lac emission: implications for IceCube neutrinos}}, \href{https://doi.org/10.1093/mnras/stv179}{\emph{\mnras} {\bfseries 448} (2015) 2412} [\href{https://arxiv.org/abs/1501.07115}{{\ttfamily 1501.07115}}].

\bibitem{2021ApJ...906..131M}
A.~{Mastichiadis} and M.~{Petropoulou}, \emph{{Hadronic X-Ray Flares from Blazars}}, \href{https://doi.org/10.3847/1538-4357/abc952}{\emph{\apj} {\bfseries 906} (2021) 131} [\href{https://arxiv.org/abs/2009.12158}{{\ttfamily 2009.12158}}].

\bibitem{1979ApJ...232..106E}
D.~{Eichler}, \emph{{High-energy neutrino astronomy: a probe of galactic nuclei?}}, \href{https://doi.org/10.1086/157269}{\emph{\apj} {\bfseries 232} (1979) 106}.

\bibitem{1981MNRAS.194....3B}
V.S.~{Berezinskii} and V.L.~{Ginzburg}, \emph{{On high-energy neutrino radiation of quasars and active galactic nuclei}}, \href{https://doi.org/10.1093/mnras/194.1.3}{\emph{\mnras} {\bfseries 194} (1981) 3}.

\bibitem{IceCube-NGC1068}
R.~{Abbasi} et~al., \emph{{Evidence for neutrino emission from the nearby active galaxy NGC 1068}}, \href{https://doi.org/10.1126/science.abg3395}{\emph{Science} {\bfseries 378} (2022) 538} [\href{https://arxiv.org/abs/2211.09972}{{\ttfamily 2211.09972}}].

\bibitem{2024arXiv240520146P}
P.~{Padovani}, E.~{Resconi}, M.~{Ajello}, C.~{Bellenghi}, S.~{Bianchi}, P.~{Blasi} et~al., \emph{{Supermassive black holes and very high-energy neutrinos: the case of NGC 1068}}, \href{https://doi.org/10.48550/arXiv.2405.20146}{\emph{arXiv e-prints} (2024) arXiv:2405.20146} [\href{https://arxiv.org/abs/2405.20146}{{\ttfamily 2405.20146}}].

\bibitem{2017ApJS..233...17R}
C.~{Ricci}, B.~{Trakhtenbrot}, M.J.~{Koss}, Y.~{Ueda}, I.~{Del Vecchio}, E.~{Treister} et~al., \emph{{BAT AGN Spectroscopic Survey. V. X-Ray Properties of the Swift/BAT 70-month AGN Catalog}}, \href{https://doi.org/10.3847/1538-4365/aa96ad}{\emph{\apjs} {\bfseries 233} (2017) 17} [\href{https://arxiv.org/abs/1709.03989}{{\ttfamily 1709.03989}}].

\bibitem{2024arXiv240607601A}
R.~{Abbasi}, M.~{Ackermann}, J.~{Adams}, S.K.~{Agarwalla}, J.A.~{Aguilar}, M.~{Ahlers} et~al., \emph{{IceCube Search for Neutrino Emission from X-ray Bright Seyfert Galaxies}}, \href{https://doi.org/10.48550/arXiv.2406.07601}{\emph{arXiv e-prints} (2024) arXiv:2406.07601} [\href{https://arxiv.org/abs/2406.07601}{{\ttfamily 2406.07601}}].

\bibitem{aguilar-ruiz_cosmic_2021}
E.~Aguilar-Ruiz, N.~Fraija, J.C.~Joshi, A.~Galvan-Gamez and J.A.~de~Diego, \emph{Cosmic rays, neutrinos and {GeV}-{TeV} gamma rays from {Starburst} {Galaxy} {NGC} 4945}, \href{https://doi.org/10.1103/PhysRevD.104.083013}{\emph{Physical Review D} {\bfseries 104} (2021) 083013}.

\bibitem{koss_population_2018}
M.J.~Koss, L.~Blecha, P.~Bernhard, C.-L.~Hung, J.R.~Lu, B.~Trakhtenbrot et~al., \emph{A population of luminous accreting black holes with hidden mergers}, \href{https://doi.org/10.1038/s41586-018-0652-7}{\emph{Nature} {\bfseries 563} (2018) 214}.

\bibitem{wang_fast_2014}
J.~Wang, E.~Nardini, G.~Fabbiano, M.~Karovska, M.~Elvis, S.~Pellegrini et~al., \emph{{FAST} {AND} {FURIOUS}: {SHOCK} {HEATED} {GAS} {AS} {THE} {ORIGIN} {OF} {SPATIALLY} {RESOLVED} {HARD} {X}-{RAY} {EMISSION} {IN} {THE} {CENTRAL} 5 kpc {OF} {THE} {GALAXY} {MERGER} {NGC} 6240}, \href{https://doi.org/10.1088/0004-637X/781/1/55}{\emph{The Astrophysical Journal} {\bfseries 781} (2014) 55}.

\bibitem{Aartsen_2021}
M.G.~Aartsen, R.~Abbasi, M.~Ackermann, J.~Adams, J.A.~Aguilar, M.~Ahlers et~al., \emph{Icecube-gen2: the window to the extreme universe}, \href{https://doi.org/10.1088/1361-6471/abbd48}{\emph{Journal of Physics G: Nuclear and Particle Physics} {\bfseries 48} (2021) 060501}.

\bibitem{Adrián-Martínez_2016}
S.~Adrián-Martínez, M.~Ageron, F.~Aharonian, S.~Aiello, A.~Albert, F.~Ameli et~al., \emph{Letter of intent for km3net 2.0}, \href{https://doi.org/10.1088/0954-3899/43/8/084001}{\emph{Journal of Physics G: Nuclear and Particle Physics} {\bfseries 43} (2016) 084001}.

\bibitem{ricci_bat_2017}
C.~Ricci, B.~Trakhtenbrot, M.J.~Koss, Y.~Ueda, I.~Del~Vecchio, E.~Treister et~al., \emph{{BAT} {AGN} {Spectroscopic} {Survey}. {V}. {X}-{Ray} {Properties} of the {Swift}/{BAT} 70-month {AGN} {Catalog}}, \href{https://doi.org/10.3847/1538-4365/aa96ad}{\emph{The Astrophysical Journal Supplement Series} {\bfseries 233} (2017) 17}.

\bibitem{2014ApJ...791...37O}
C.A.~{Onken}, M.~{Valluri}, J.S.~{Brown}, P.J.~{McGregor}, B.M.~{Peterson}, M.C.~{Bentz} et~al., \emph{{The Black Hole Mass of NGC 4151. II. Stellar Dynamical Measurement from Near-infrared Integral Field Spectroscopy}}, \href{https://doi.org/10.1088/0004-637X/791/1/37}{\emph{\apj} {\bfseries 791} (2014) 37} [\href{https://arxiv.org/abs/1406.6735}{{\ttfamily 1406.6735}}].

\bibitem{2008ApJS..174...31H}
E.K.S.~{Hicks} and M.A.~{Malkan}, \emph{{Circumnuclear Gas in Seyfert 1 Galaxies: Morphology, Kinematics, and Direct Measurement of Black Hole Masses}}, \href{https://doi.org/10.1086/521650}{\emph{\apjs} {\bfseries 174} (2008) 31} [\href{https://arxiv.org/abs/0707.0611}{{\ttfamily 0707.0611}}].

\bibitem{2018ApJ...866..133D}
G.~{De Rosa}, M.M.~{Fausnaugh}, C.J.~{Grier}, B.M.~{Peterson}, K.D.~{Denney}, K.~{Horne} et~al., \emph{{Velocity-resolved Reverberation Mapping of Five Bright Seyfert 1 Galaxies}}, \href{https://doi.org/10.3847/1538-4357/aadd11}{\emph{\apj} {\bfseries 866} (2018) 133} [\href{https://arxiv.org/abs/1807.04784}{{\ttfamily 1807.04784}}].

\bibitem{2020ApJ...902...26Y}
W.~{Yuan}, M.M.~{Fausnaugh}, S.L.~{Hoffmann}, L.M.~{Macri}, B.M.~{Peterson}, A.G.~{Riess} et~al., \emph{{The Cepheid Distance to the Seyfert 1 Galaxy NGC 4151}}, \href{https://doi.org/10.3847/1538-4357/abb377}{\emph{\apj} {\bfseries 902} (2020) 26} [\href{https://arxiv.org/abs/2007.07888}{{\ttfamily 2007.07888}}].

\bibitem{2017ApJ...850...74K}
M.~{Koss}, B.~{Trakhtenbrot}, C.~{Ricci}, I.~{Lamperti}, K.~{Oh}, S.~{Berney} et~al., \emph{{BAT AGN Spectroscopic Survey. I. Spectral Measurements, Derived Quantities, and AGN Demographics}}, \href{https://doi.org/10.3847/1538-4357/aa8ec9}{\emph{\apj} {\bfseries 850} (2017) 74} [\href{https://arxiv.org/abs/1707.08123}{{\ttfamily 1707.08123}}].

\bibitem{greenhill_distribution_nodate}
L.J.~Greenhill, J.M.~Moran and J.R.~Herrnstein, \emph{{THE} {DISTRIBUTION} {OF} {H2O} {MASER} {EMISSION} {IN} {THE} {NUCLEUS} {OF} {NGC} 4945}, .

\bibitem{karachentsev_hubble_2007}
I.D.~Karachentsev, R.B.~Tully, A.~Dolphin, M.~Sharina, L.~Makarova, D.~Makarov et~al., \emph{The {Hubble} flow around the {CenA} / {M83} galaxy complex}, \href{https://doi.org/10.1086/510125}{\emph{The Astronomical Journal} {\bfseries 133} (2007) 504}.

\bibitem{koss_bass_2022}
M.J.~Koss, C.~Ricci, B.~Trakhtenbrot, K.~Oh, J.S.~Den~Brok, J.E.~Mejía-Restrepo et~al., \emph{{BASS}. {XXII}. {The} {BASS} {DR2} {AGN} {Catalog} and {Data}}, \href{https://doi.org/10.3847/1538-4365/ac6c05}{\emph{The Astrophysical Journal Supplement Series} {\bfseries 261} (2022) 2}.

\bibitem{ajello_disentangling_2023}
M.~Ajello, K.~Murase and A.~McDaniel, \emph{Disentangling the {Hadronic} {Components} in {NGC} 1068}, \href{https://doi.org/10.3847/2041-8213/acf296}{\emph{The Astrophysical Journal Letters} {\bfseries 954} (2023) L49}.

\bibitem{murase_sub-gev_2024}
K.~Murase, C.M.~Karwin, S.S.~Kimura, M.~Ajello and S.~Buson, \emph{Sub-{GeV} {Gamma} {Rays} from {Nearby} {Seyfert} {Galaxies} and {Implications} for {Coronal} {Neutrino} {Emission}}, \href{https://doi.org/10.3847/2041-8213/ad19c5}{\emph{The Astrophysical Journal Letters} {\bfseries 961} (2024) L34}.

\bibitem{abbasi_search_2024}
R.~Abbasi, M.~Ackermann, J.~Adams, S.K.~Agarwalla, J.A.~Aguilar, M.~Ahlers et~al., \emph{Search for neutrino emission from hard {X}-ray {AGN} with {IceCube}},  June, 2024.

\bibitem{trakhtenbrot_bat_2017}
B.~Trakhtenbrot, C.~Ricci, M.J.~Koss, K.~Schawinski, R.~Mushotzky, Y.~Ueda et~al., \emph{{BAT} {AGN} {Spectroscopic} {Survey} ({BASS}) – {VI}. {The} $\gamma${X}–{L}/{LEdd} relation}, \href{https://doi.org/10.1093/mnras/stx1117}{\emph{Monthly Notices of the Royal Astronomical Society} {\bfseries 470} (2017) 800}.

\bibitem{Lightman}
A.P.L.~George B.~Rybicki, \emph{Radiative processes in astrophysics}, Haruard-Smithsonian Center for Astrophysics.

\bibitem{fiorillo_magnetized_2024}
D.F.G.~Fiorillo, L.~Comisso, E.~Peretti, M.~Petropoulou and L.~Sironi, \emph{A magnetized strongly turbulent corona as the source of neutrinos from {NGC} 1068}, \href{https://doi.org/10.3847/1538-4357/ad7021}{\emph{The Astrophysical Journal} {\bfseries 974} (2024) 75}.

\bibitem{waxman_high_1997}
E.~Waxman and J.~Bahcall, \emph{High {Energy} {Neutrinos} from {Cosmological} {Gamma}-{Ray} {Burst} {Fireballs}}, \href{https://doi.org/10.1103/PhysRevLett.78.2292}{\emph{Physical Review Letters} {\bfseries 78} (1997) 2292}.

\bibitem{petropoulou_implications_2014}
M.~Petropoulou, D.~Giannios and S.~Dimitrakoudis, \emph{Implications of a {PeV} neutrino spectral cutoff in {GRB} models}, \href{https://doi.org/10.1093/mnras/stu1757}{\emph{Monthly Notices of the Royal Astronomical Society} {\bfseries 445} (2014) 570}.

\bibitem{florou_marginally_2021}
I.~Florou, M.~Petropoulou and A.~Mastichiadis, \emph{A marginally fast-cooling proton–synchrotron model for prompt {GRBs}}, \href{https://doi.org/10.1093/mnras/stab1285}{\emph{Monthly Notices of the Royal Astronomical Society} {\bfseries 505} (2021) 1367}.

\bibitem{2024A&A...684L..21P}
P.~{Padovani}, R.~{Gilli}, E.~{Resconi}, C.~{Bellenghi} and F.~{Henningsen}, \emph{{The neutrino background from non-jetted active galactic nuclei}}, \href{https://doi.org/10.1051/0004-6361/202450025}{\emph{\aap} {\bfseries 684} (2024) L21} [\href{https://arxiv.org/abs/2404.05690}{{\ttfamily 2404.05690}}].

\bibitem{2017MNRAS.470.1881P}
M.~{Petropoulou}, S.~{Coenders}, G.~{Vasilopoulos}, A.~{Kamble} and L.~{Sironi}, \emph{{Point-source and diffuse high-energy neutrino emission from Type IIn supernovae}}, \href{https://doi.org/10.1093/mnras/stx1251}{\emph{\mnras} {\bfseries 470} (2017) 1881} [\href{https://arxiv.org/abs/1705.06752}{{\ttfamily 1705.06752}}].

\bibitem{1990MNRAS.245..453C}
P.S.~{Coppi} and R.D.~{Blandford}, \emph{{Reaction rates and energy distributions for elementary processes in relativistic pair plasmas}}, \href{https://doi.org/10.1093/mnras/245.3.453}{\emph{\mnras} {\bfseries 245} (1990) 453}.

\bibitem{1992ApJ...400..181C}
M.J.~{Chodorowski}, A.A.~{Zdziarski} and M.~{Sikora}, \emph{{Reaction Rate and Energy-Loss Rate for Photopair Production by Relativistic Nuclei}}, \href{https://doi.org/10.1086/171984}{\emph{\apj} {\bfseries 400} (1992) 181}.

\bibitem{kelner_energy_2008}
S.R.~Kelner and F.A.~Aharonian, \emph{Energy spectra of gamma-rays, electrons and neutrinos produced at interactions of relativistic protons with low energy radiation}, \href{https://doi.org/10.1103/PhysRevD.78.034013}{\emph{Physical Review D} {\bfseries 78} (2008) 034013}.

\end{thebibliography}\endgroup

\appendix
\section{Analytical expression for $ L_{\rm {\nu}+\bar{\nu}}$}\label{app:Lnu-bolo}
We compute the bolometric neutrino luminosity in the optically thin regime by integrating $ L_{\rm \nu+\bar{\nu}}(E_{\rm \nu})$ given in Eq.~\ref{eq:Lnu}. 

We first examine the case with $ E_{\rm \nu, \rm br} \simeq 5~\sigma_{\rm p,5}{\rm TeV} \le E_{\rm \nu,*}\simeq 46.5~{\rm TeV}$, that holds for $ \sigma_{\rm p} \lesssim 10^6$. Then, there are three energy ranges of interest: (i) $ E_{\rm \nu} \le E_{\rm \nu, \rm br}$, (ii) $ E_{\rm \nu, \rm br} < E_{\rm \nu} \le E_{\rm \nu,*}$, and (iii) $ E_{\rm \nu, *} < E_{\rm \nu} \le E_{\rm \nu, \max} \simeq 500~{\rm TeV}~\gamma_{\rm p,\max, 8}$. We obtain,

\begin{eqnarray}
L_{\rm \nu+\bar{\nu}}^{\rm (thin)} = A\frac{E_{\rm \nu, \rm br}}{25~\rm TeV}
\left [  \frac{1}{2}\left(1-\frac{E_{\rm \nu, \min}^2}{E_{\rm \nu, \rm br}^2} \right) + \ln \left( \frac{E_{\rm \nu, *}}{E_{\rm \nu, \rm br}}\right) +
1-\frac{E_{\rm \nu, *}}{E_{\rm \nu, \max}}.
\right]    
\end{eqnarray}
where $E_{\rm \nu, \min} \ll E_{\rm \nu, \rm br}$ is the energy of the less energetic neutrinos that are produced by protons interacting near the threshold with coronal photons of energy $E_{\rm X, max}$,  and 
\begin{gather}
A = \frac{18}{16}\frac{\eta_{\rm p}}{\eta_{\rm X}} L_{\rm X} \frac{\lambda_{\rm X, Edd,-2}}{\tilde{R}}.
\end{gather}
For proton magnetizations $ \sigma_{\rm p} > 10^6$, the energy regimes of interest for the integration become (i) $ E_{\rm \nu} \le E_{\rm \nu, *}$, (ii) $ E_{\rm \nu,*} \le E_{\rm \nu} < E_{\rm \nu, \rm br}$, and (iii) $ E_{\rm \nu, \rm br} < E_{\rm \nu} \le E_{\rm \nu, \max}$, and the bolometric luminosity reads
\begin{eqnarray}
L_{\rm \nu+\bar{\nu}}^{\rm (thin)} =  A\frac{E_{\rm \nu, *}}{25~\rm TeV}
\left [
\frac{E_{\rm \nu, *}}{2E_{\rm \nu, \rm br}} \left(1- \frac{E_{\rm \nu, \min}^2}{E_{\rm \nu, *}^2} \right) + 1- \frac{E_{\rm \nu, *}}{E_{\rm \nu, \rm br}} + 1 -\frac{E_{\rm \nu, \rm br}}{E_{\rm \nu, \max}}.
\right]
\end{eqnarray}
In the optically thick regime the result is simply written as 
\begin{equation}
\label{eq:Lnu_LX_thick}
L_{\rm \nu+\bar{\nu}}^{\rm (thick)} \approx \frac{3}{8}   \frac{\eta_{\rm p}}{\eta_{\rm X}}L_{\rm X}
\end{equation}
where we assumed $ E_{\rm \nu, \min} \ll E_{\rm \nu, \rm br} \ll E_{\rm \nu, \max}$.

The expression derived above is valid when the timescale of the photopion production, $t_{\rm p \gamma}$ is significantly shorter than the timescales of all the other hadronic interactions, such as proton synchrotron or Bethe-Heitler pair production. If the aforementioned condition is not satisfied, the proton population loses a fraction of its energy due to other interactions apart from photopion production. As a result the amount of energy transferred to neutrinos is reduced in comparison to cases where $t_{\rm p \gamma} \ll t_{\rm p,syn}$ and $t_{\rm p \gamma} \ll  t_{\rm BH}$.

\section{Other channels for pair production}\label{App:channels} 
\subsection{The photomeson (p$ \gamma$) production channel} \label{App:pgamma}
We consider the production of neutral pions ($ \pi^0$) through interactions of protons with X-ray photons. Neutral pions of energy $ E_{\rm \pi^0} = k_{\rm p\gamma}E_{\rm p} \approx 0.2 E_{\rm p}$ will instantaneously decay into two $ \gamma$-ray photons.  The average $ \gamma$-ray photon energy (in the lab frame) is taken to be half of that of the parent pion, 
    \begin{gather}
        E_{\rm \gamma}= \frac{1}{2} E_{\rm \pi^0} \approx 0.1 E_{\rm p, \rm br} \approx 2.5~{\rm TeV} \left(\frac{E_{\rm p, \rm br}}{25~\rm TeV}\right).
    \end{gather}
The TeV photons from pion decay may interact with low-energy photons, i.e., $ E E_\gamma \ge 2 (m_{\rm e} c^2)^2 \Rightarrow E\ge 0.2~{\rm eV} \left(\frac{E_{\rm p, \rm br}}{25~\rm TeV}\right)^{-1}$, through photon-photon annihilation to produce electron-positron pairs. The interactions of TeV photons with X-ray coronal photons ($ E_X\ge 0.1$~keV) will take place far from the energy threshold of $ \gamma \gamma$ pair production, where the cross section will be about three orders of magnitude lower than its maximum value (close to the threshold)~\cite{1990MNRAS.245..453C}. Nonetheless, photons produced by the secondary pairs themselves via synchrotron radiation will provide a sufficient number of low-energy targets in order for the TeV photons to interact close to the threshold. This will become clear 
by our numerical results that account for all photon populations in the corona  (see section~\ref{sec:res}).

We therefore assume that all the luminosity injected into pions is transferred to $ \gamma$-ray photons, which in turn will be attenuated, thus transferring their luminosity into secondary pairs. 
We may express the pair energy injection rate as 
\begin{gather}
   L^{p\gamma}_{\rm \pm} \approx  L_{\rm \pi^0} \approx \frac{1}{2} f_{\rm p\gamma}  E_{\rm p, \rm br}L_{\rm p} (E_{\rm p, \rm br})
\end{gather}
where $f_{\rm p\gamma}$ is defined in Eq.~\ref{eq:f_pg}. 
    
Regarding charged pion production, we also assume that the pion is created with 20 per cent the energy of the parent proton. The pion will decay, resulting in the creation of 3 neutrinos and one lepton. If the pion energy is equally distributed to all the secondary particles, then the lepton energy upon production reads, 
\begin{gather}
    E_{\rm \pm}^{p\gamma\rightarrow \pi^\pm \rightarrow ee} \approx \frac{1}{4} E_{\rm \pi^\pm} = 0.05 E_{\rm p, \rm br}.
    \label{eq:EE_pg}
\end{gather}
Summing up the contributions of the neutral pion and charged pion decays, and using Eq.~\ref{eq:n_pm} we estimate the secondary pair number density as
\begin{gather}
n^{ p\gamma}_{\rm \pm} \approx 0.13 \cdot 10^7~{\rm cm}^{-3}\frac{\eta'_p}{\eta_X}\left(\frac{E_{\rm p, \rm br}}{25~\rm TeV}\right)^{-1}\frac{\lambda_{\rm X, Edd,-2}^2}{\tilde{R}^2 L_{\rm X,43}}\min  \left \{ 1, 0.3 \frac{\lambda_{\rm X, Edd, -2}} {\tilde{R}}\frac{\min \left(E_{\rm p, br}, E_{\rm p,*}\right) }{25\, \rm TeV} \right\}. \label{n_ee_pg}
\end{gather}
In the optically thick limit, we find $ n^{ p\gamma}_{\rm \pm} \propto \lambda_{\rm X, Edd}^2 L_{\rm X}^{-1} \tilde{R}^{-2} \propto L_{\rm X} R^{-2}$ whereas in the optically thin limit the dependence on coronal parameters is stronger, $ n^{p\gamma}_{\rm \pm} \propto \lambda_{\rm X, Edd}^3 L_{\rm X}^{-1} \tilde{R}^{-3} \propto L_{\rm X}^2 R^{-3}$. 

\subsection{Proton-photon (Bethe-Heitler) pair production} \label{App:BH}
Another hadronic channel for pair production is proton-photon (Bethe-Heitler) pair production. The bolometric luminosity transferred from the proton population to the secondary pair population can be written as
        \begin{gather}
            L^{BH}_{\rm \pm}=f_{\rm BH} E_{\rm p, \rm br}L_{\rm p} (E_{\rm p, \rm br}) \label{Eq:L_BH}
        \end{gather}
        where $f_{\rm BH}$ denotes the efficiency of this process, which is written as 
        \begin{gather}
            f_{\rm BH}=\min \left \{ 1, \hat{\sigma}_{\rm BH} n_t R \right \}
            \label{Eq:f_BH}
        \end{gather}
where $ \hat{\sigma}_{\rm BH} \simeq 8\times10^{-31}$~cm$^2$ is the maximum effective cross section accounting for the inelasticity of the interaction \cite{1992ApJ...400..181C}, $ n_t = 3 L_t/\left(4\pi R^2 c \epsilon_t\right)$ is the number density of target photons for Bethe-Heitler, and $ \epsilon_t$ is the target photon energy. The latter is expressed as $ \epsilon_t \simeq 16 m_{\rm e} c^2/\gamma_{\rm p} \approx 0.32~{\rm keV} (25~{\rm TeV}/ E_{\rm p, \rm br})$ assuming interactions close to the threshold~\cite{mastichiadis_spectral_2005, karavola_BH_2024}. 

We also assume that the created pairs have a Lorentz factor equal to the one of the parent proton (this Lorentz factor corresponds to the peak of the energy distribution of injected pairs as shown in \cite{karavola_BH_2024}),
\begin{gather}
    E_{\rm \pm}^{BH}=E_{\rm p, \rm br} \frac{m_{\rm e}}{m_{\rm p}}. \label{eq:E_BH}
\end{gather}

Substitution of expressions \ref{Eq:L_BH}, \ref{Eq:f_BH} and \ref{eq:E_BH} into Eq.~\ref{eq:n_pm} yields
    \begin{gather}
        n_{\rm \pm}^{BH}=1.2 \cdot 10^8~{\rm cm^{-3}} \frac{\eta_p^{\prime}}{\eta_X} 
        \left(\frac{E_{\rm p, \rm br}}{25~\rm TeV} \right)^{-1} \frac{\lambda_{\rm X, Edd,-2}^2}{\tilde{R}^2 L_{\rm X,43}} \min \left \{ 1, 0.1 \frac{\lambda_{\rm X, Edd,-2}}{\tilde{R}} \left(\frac{E_{\rm p, \rm br}}{25~\rm TeV}\right) \right \} \label{n_ee_BH-new}
    \end{gather}

\section{Proton distribution with a post-break slope of 2} \label{App:slope_2}

In this work we describe the proton differential number distribution as a broken power law with a break at $ \gamma_{\rm p}=\sigma_{\rm p}$, with $ \sigma_{\rm p}$ being the proton magnetization parameter (see Eq. \ref{eq:dNdEdt}). The pre-break slope is fixed at $-1$, while the post-break slope value we used was set to $-3$ (or $s_{\rm p}=3$ equivalently). We compute the electromagnetic and neutrino spectra for the same parameters as those used in figure \ref{fig:spectr_neutr_prot}, with the only change being the post-break slope of the proton differential number distribution, which is, now, $s_{\rm p}=2$. 

\begin{figure}[h]
\centering        \includegraphics[width=0.45\textwidth]{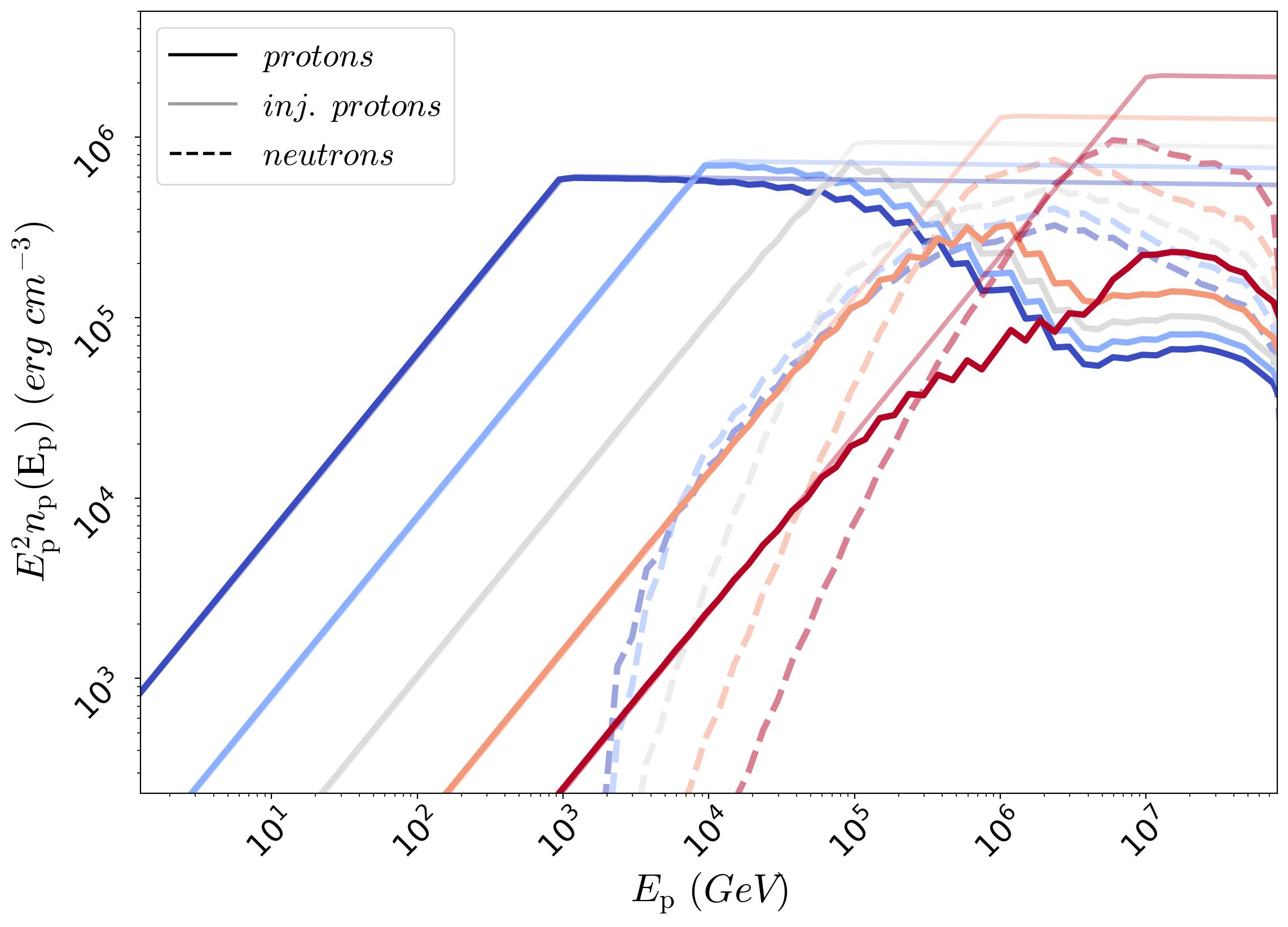}
        \includegraphics[width=0.495\textwidth]{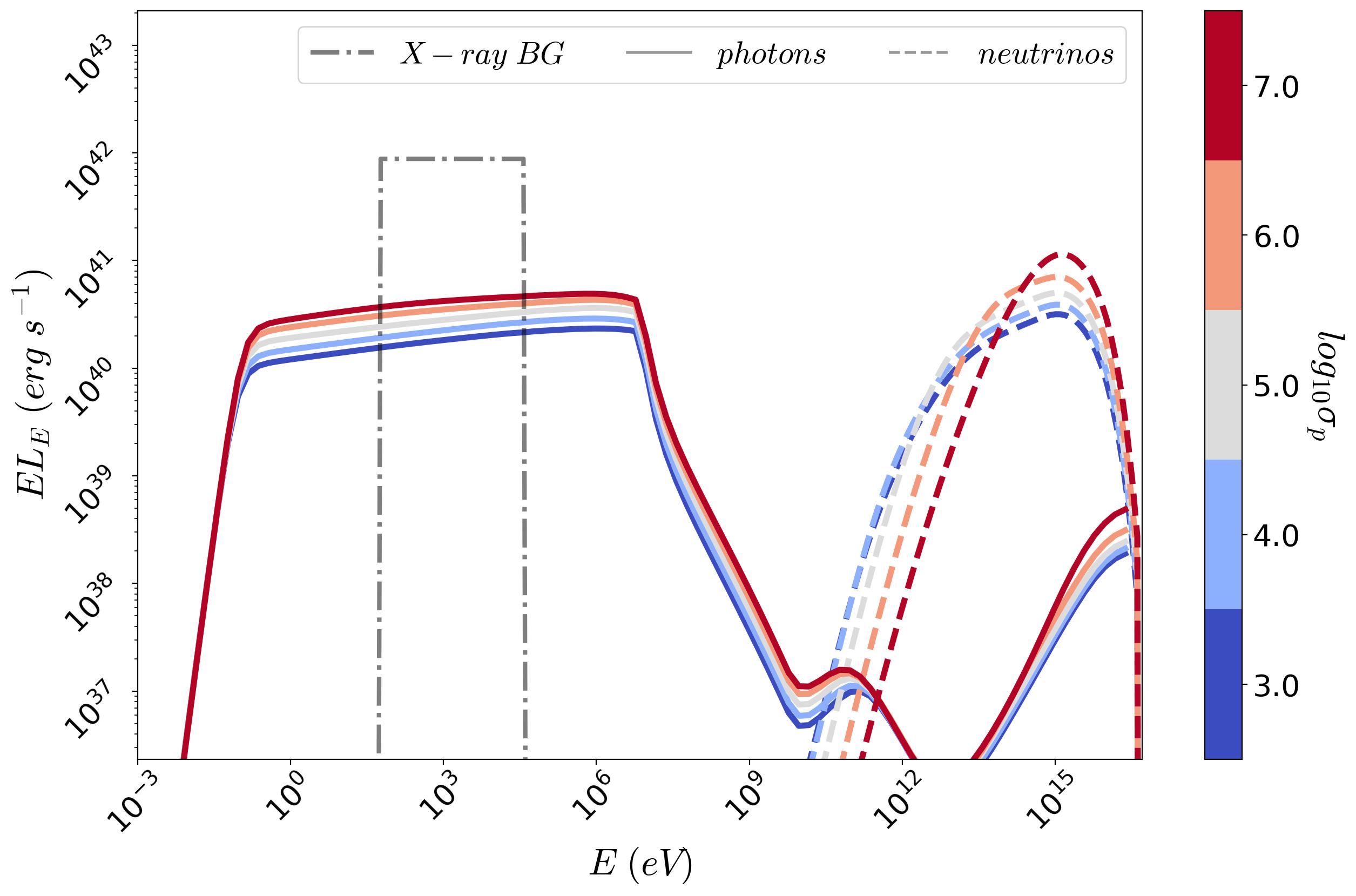}
        \caption{Same as figure \ref{fig:spectr_neutr_prot} but for a post-break slope of the proton distribution of $s_{\rm p}=2$.}
        \label{fig:sp_2}
\end{figure}

Our results are presented in figure \ref{fig:sp_2}. We observe that the cascade spectrum does not highly depend on $ \sigma_{\rm p}$ value  when $s_{\rm p}=2$. In this case the energy is equally distributed (in logarithm) to protons with $ \gamma_{\rm p} \ge \sigma_{\rm p}$.  Combining the latter with the almost constant value of the photopion effective cross section \cite{kelner_energy_2008}, the protons above the break of the distribution can find and interact efficiently with the luminous coronal photons. The almost universal photon spectrum, especially at energies where $ \gamma \gamma$ annihilation is most important as a source of pairs ($\sim 10$~MeV, see also section~\ref{sec:analytic-pairs}), suggests that the secondary pair density will also not depend strongly on $ \sigma_{\rm p}$. 

Furthermore, we highlight that the neutrino spectrum is not significantly affected by the change of $ \sigma_{\rm p}$. We notice that the peak neutrino luminosity changes only by a factor of 3 when $ \sigma_{\rm p}$ varies by 4 orders of magnitude. Most importantly, the peak energy of the neutrino spectrum, also, remains unchanged when we vary $ \sigma_{\rm p}$, as it is controlled by the highest energy protons of the distribution. 
Therefore, if the post-break slope of the proton distribution is $ s_{\rm p}=2$, both the neutrino peak energy and the neutrino post-break slope are insensitive to $ \sigma_{\rm p}$, as opposed to the $ s_{\rm p}=3$ case (see figure \ref{fig:spectr_neutr_prot}). This is an important qualitative difference that could be used to distinguish between possible post-break slopes of the accelerated proton population.

Finally, we note that changing $ s_{\rm p}$ will not significantly alter the behaviour of the system in the $ n_{\rm \pm}-\lambda_{\rm X, Edd}$ or $ L_{\rm \nu + \bar{\nu}}/L_{\rm X}-\lambda_{\rm X, Edd}$ parameter spaces, since both quantities depend on the broadband corona luminosity (see Eqs. \ref{eq:nuLnu} and \ref{n_ee_gg-new} and figure \ref{fig:ne_max}).

\section{$ \gamma-\gamma$ optical depth} \label{App:gg}

In section \ref{sec:res} we presented the numerical results of our model. In particular, in the right panel of figure \ref{fig:spectr_neutr_prot}, we show the cascade electromagnetic spectrum of the coronal region, noting that its morphology is similar for all $ \sigma_{\rm p}$ values, as a result of $ \gamma \gamma$ annihilation for photons with energies $\geq$ 10 MeV. In order to validate this we numerically calculate the optical depth of the $ \gamma \gamma$ annihilation process, as
\begin{gather}
  \tau_{\rm \gamma \gamma}(\epsilon_\gamma ) \approx 0.625 \sigma_{\rm T} R \int_{\rm 2/\epsilon_{\rm \gamma}} \d \epsilon^\prime n(\epsilon^\prime) \frac{(\epsilon_{\rm \gamma} \epsilon^\prime)^{2}-1}{(\epsilon_{\rm \gamma} \epsilon^\prime)^3} \ln(\epsilon_{\rm \gamma} \epsilon^\prime) 
    \label{tau_gg_mono}
\end{gather}
where we used the approximate $ \gamma \gamma $ cross section from Ref.~\cite{1990MNRAS.245..453C} and with $ \epsilon_{\rm \gamma}=E_{\rm \gamma}/(m_{\rm e} c^2)$, $ R$ being the radius of the coronal region, $ \sigma_{\rm T}$ being the Thomson cross section, and $ n(\epsilon)$ representing the differential number distribution of target photons. 

\begin{figure}[h!]
    \centering
    \includegraphics[width=0.48\linewidth]{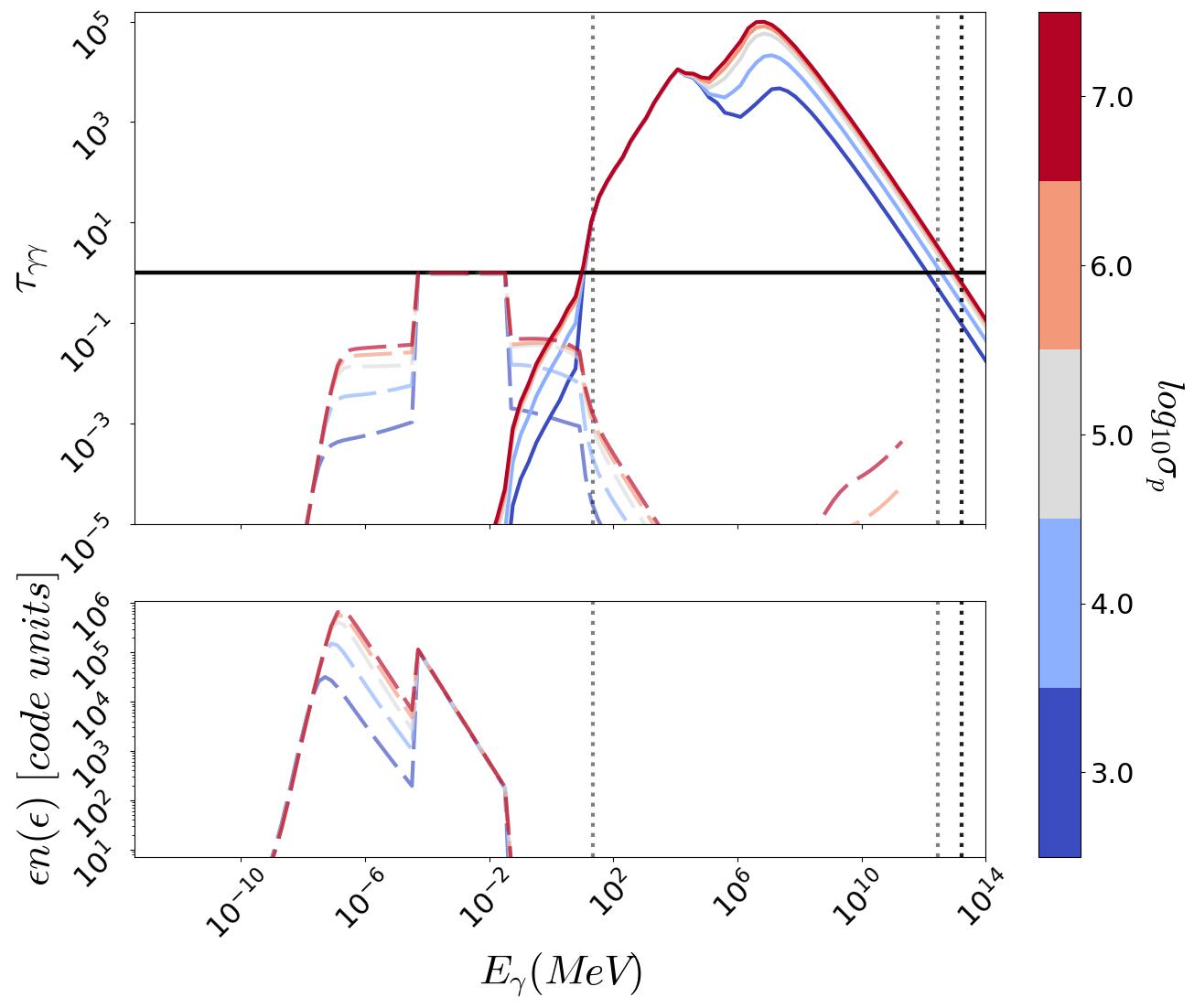}
    \includegraphics[width=0.495\linewidth]{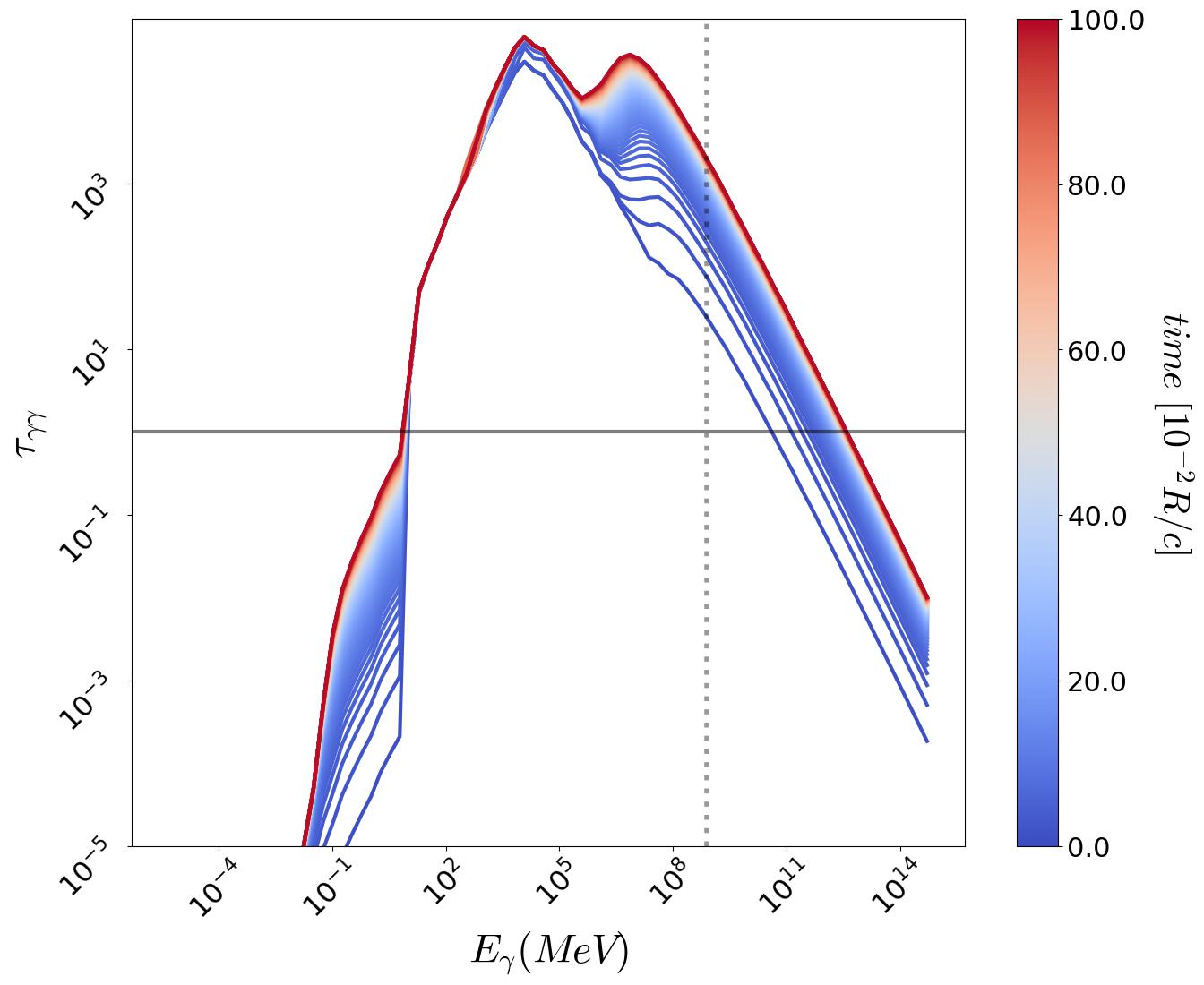}
    \caption{\textit{Top panel:} Solid lines represent the optical depth for $ \gamma \gamma$ annihilation process as a function on the high energy photon that is absorbed for the cases of figure \ref{fig:spectr_neutr_prot}. Colorbar shows the proton magnetization value, while dotted lines represent the cascade electromagnetic spectrum for each of the cases, normalized to its maximum. Vertical dotted line in both panels represent $ \tau_{\rm \gamma \gamma} =1$ and we see that photons of higher energies are significantly attenuated. \textit{Bottom panel:} Dashed lines show the number density of the photon distribution acting as targets for $ \gamma \gamma$, for the same cases as the ones in the top panel. We note how the morphology of the target photon number distribution reflects in the shape of $ \tau_{\rm \gamma \gamma}$. \textit{Right panel:} Time evolution of the optical depth for the case of $ \sigma_{\rm p}=10^5$, presented in the left panels. The time span shown is one light-crossing time with a step between the snapshots of 0.01 $ R/c$.}
    \label{fig:tau_gg}
\end{figure}

The results of the $ \tau_{\rm \gamma \gamma}$ calculation for the cases of figure \ref{fig:spectr_neutr_prot}, are shown in figure \ref{fig:tau_gg}. 
The solid lines in the top left panel represent the optical depth as a function of the energy of photons that are annihilated. In the same panel, the dashed lines show the cascade electromagnetic spectrum for each of the cases (in a $ \epsilon^2 n(\epsilon)$ versus $ \epsilon$ representation), while the dashed lines in the bottom right panel refer to the number distribution of the photon population in the corona. Different colors correspond to different $ \sigma_{\rm p}$ values (see color bar). We note that the shape of the photon number distribution is reflected to the optical depth with the peak at $ 10^{-10}$ MeV being responsible for the very high opacity value at $ 10^{10}$ MeV and thus the absorption of all the luminosity produced at such energies (mostly due to photomeson -- see appendix~\ref{App:channels}). Therefore, the low-energy photons from the electromagnetic cascade can attenuate the TeV to PeV $ \gamma$ rays, even in the absence of other ambient dense low-energy photons (e.g. from the disk).
We, also, see a second peak in the photon distribution at energies of $ \sim 10^{-3}$ MeV that corresponds to the minimum energy photons of the X-ray corona. This secondary peak in the target photon number density is reflected to $ \tau_{\rm \gamma \gamma}$ at energies $ \sim 10^{3}$ MeV. Furthermore, the vertical dotted line indicates the photon energy where the system goes from the optically thin to the optically thick regime. The aforementioned transition happens at $ \sim 10$ MeV and we notice that photons with higher energies are attenuated almost completely.

Moreover, in the right panel of figure \ref{fig:tau_gg}, we show the time evolution of the optical depth for one of the cases displayed on the left panel, the one with $ \sigma_{\rm p}=10^5$. Results are shown for a time interval of one light crossing time, $R/c$, since this is long enough for a steady state to be reached. We see that initially the system is optically thin to photons with PeV energy values, as there are no photons below 100~eV initially present in the corona. However, as time progresses and photopion production contributes to the TeV-PeV $ \gamma$-ray emission, secondary pairs radiate low energy photons acting as $ \gamma \gamma$ targets to the former, and the corona becomes optically thick to very high-energy $ \gamma$ rays within one light crossing time.

\section{Effects of meson synchrotron cooling}\label{App:cooling} 

\begin{figure}[h!]
    \centering \includegraphics[width=0.9\linewidth]{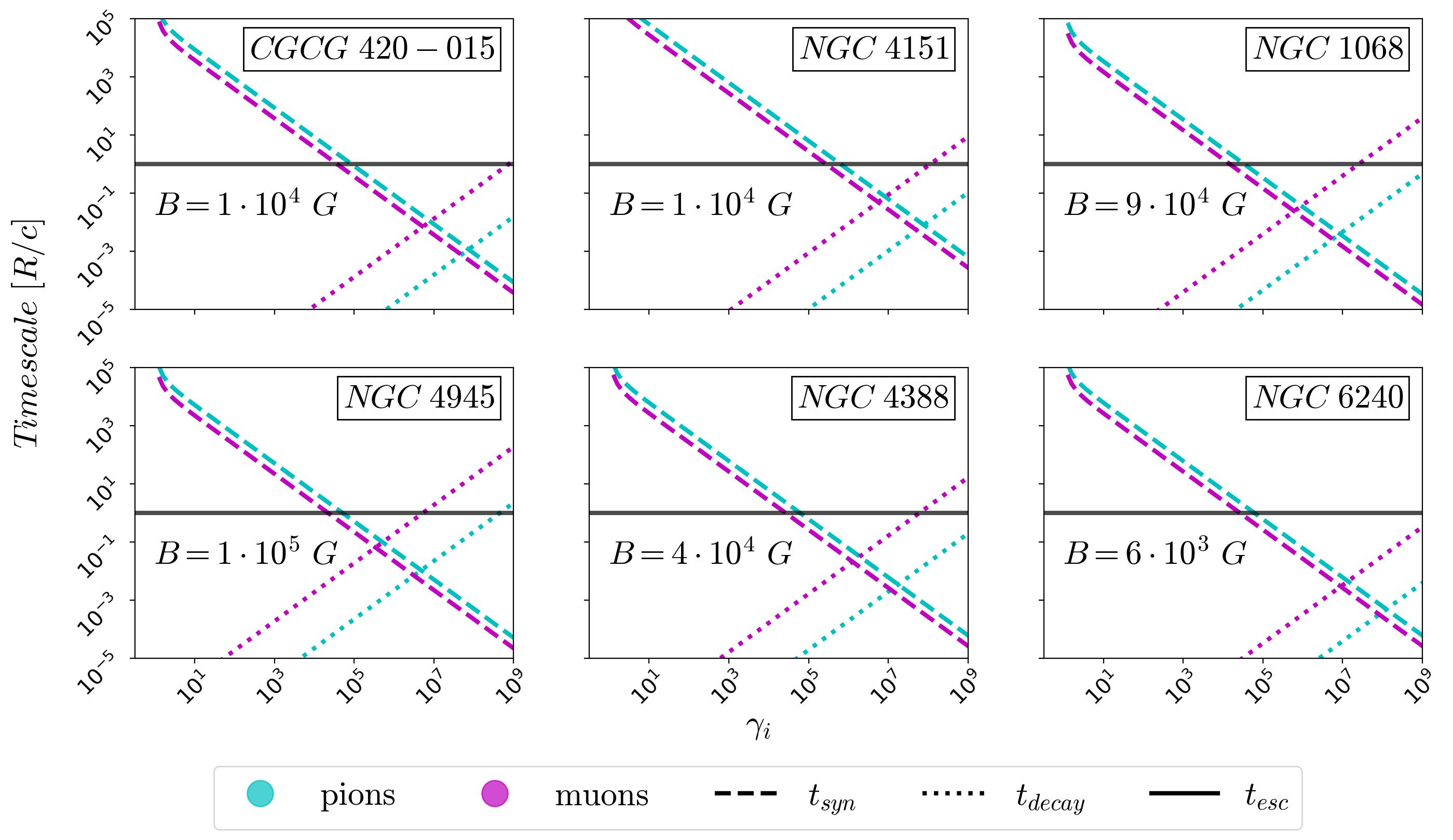}
    \caption{Synchrotron cooling and decay timescales (dashed and dotted lines respectively) for charged pions (cyan) and muons (magenta) in $R/c$ units for the Seyfert galaxies in figure \ref{fig:seyf_gal_neutr}. Black solid line shows the escape timescale.}
    \label{fig:heavy_times}
\end{figure}

Photohadronic interactions produce secondary particles, such as pions and muons that are unstable and decay into lighter particles and/or photons (see diagram in Sec.~\ref{sec:analytic-pairs}). In highly magnetized environments, such as the ones discussed in the present work, the aforementioned heavy particles might be able to radiate a large portion of their kinetic energy before they decay. If this is the case, then their radiation can contribute to the electromagnetic cascade spectrum and, more importantly, less energy will be transferred to neutrinos. In order to examine whether synchrotron radiation of pions and muons is important in the scope of our work, we compare the timescales of decay and synchrotron losses for these particles; see figure~\ref{fig:heavy_times}.

% \begin{gather}
%     t_{i, syn}=\frac{E_i}{- \frac{d E_i}{d t}|_{syn }} = \frac{\gamma_i m_i c^2}{\frac{4}{3}\sigma_T c \left( \frac{m_e}{m_i} \right)^2 \gamma_i^2 \frac{B^2}{8 \pi}}
% \end{gather}

\begin{figure}[h!]
    \centering
    \includegraphics[width=0.9\linewidth]{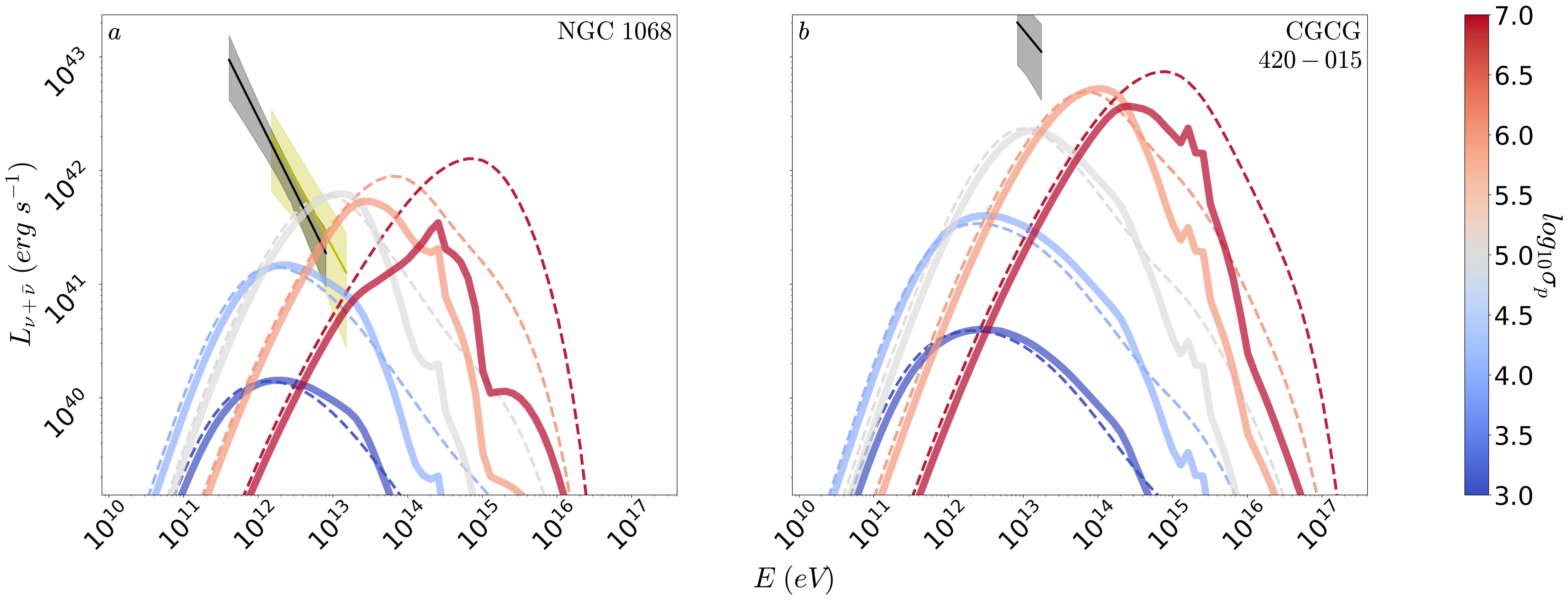}
    \caption{Numerical results of all-flavor neutrino spectra for NGC~1068 (panel a) and CGCG~420-015 (panel b) computed for a range of $ \sigma_{\rm p}$ values (see color bar). The dashed lines represent the results without taking into account the synchrotron cooling for pions and muons, as opposed to the solid lines.}
    \label{fig:seyf_gal_neutr_heavy}
\end{figure}

In figure \ref{fig:seyf_gal_neutr_heavy} we show the neutrino energy distributions for two sources, NGC~1068 and CGCG~420-015, with indicative magnetic field strengths of $10^4$G and $10^5$G respectively-- see also figure \ref{fig:heavy_times}. Dashed lines represent the results without cooling for the heavier secondary populations, such as pions and muons (see also figure~ \ref{fig:seyf_gal_neutr}), while solid lines show the results with pion and muon synchrotron cooling taken into account. One can notice that synchrotron cooling of mesons alters both the shape of the neutrino distribution and the luminosity and position of its maximum for sources with higher magnetic fields like NGC~1068 (for which our model yields $9\cdot 10^4$ G) and high $\sigma_p$ values. However, if the magnetization is lower, for example $\sigma_p=10^5$ (such as that needed to describe the IceCube data for NGC~1068), the luminosity and position of the distribution peak are not significantly affected, even though the high-energy tail of the spectrum becomes steeper, a result that better matches the observations. For sources with weaker magnetic fields like  CGCG~420-015, the effects of pion and muon cooling become relevant for higher magnetization values, e.g. $\sigma_p > 10^6$. The aforementioned importance of the magnetic field can also be derived by figure \ref{fig:heavy_times} in which we see the synchrotron timescale becoming comparable to the decay one for pion/muon Lorentz factor values $\gamma_i \sim 5 \cdot 10^5$ (with $\gamma_i=\kappa_{p\gamma, meson} m_p/m_i \gamma_p =0.2 m_p/m_i \gamma_p \propto \sigma_p$, with $m_i$ being the particle mass and $\kappa_{p\gamma, meson}$ being the fraction of the proton energy transferred to the pion/muon), as the magnetic field increases, see NGC~4945. On the contrary for NGC~6240 that has a weaker magnetic field, the dashed and dotted lines do not cross with each other for $\gamma_i \lesssim 10^7$.

\end{document}